\documentclass[10pt,english,a4paper,aps,showpacs,showkeys,amsmath,amssymb,nofootinbib,pre,floatfix,preprint]{revtex4-1}
\usepackage[T1]{fontenc}
\usepackage[utf8]{inputenc}
\usepackage{verbatim}
\usepackage{amsmath}
\usepackage{graphicx}
\usepackage{amssymb}
\usepackage{esint}

\makeatletter

\@ifundefined{textcolor}{}
{%
 \definecolor{BLACK}{gray}{0}
 \definecolor{WHITE}{gray}{1}
 \definecolor{RED}{rgb}{1,0,0}
 \definecolor{GREEN}{rgb}{0,1,0}
 \definecolor{BLUE}{rgb}{0,0,1}
 \definecolor{CYAN}{cmyk}{1,0,0,0}
 \definecolor{MAGENTA}{cmyk}{0,1,0,0}
 \definecolor{YELLOW}{cmyk}{0,0,1,0}
 }

\usepackage{amsfonts}

\usepackage{xcolor}
\usepackage{slashed}
\usepackage{feynmp}

\DeclareSymbolFont{rsfs}{U}{rsfs}{m}{n}
\DeclareSymbolFontAlphabet{\mathrsfs}{rsfs}

\newcommand{\bbeta}{\boldsymbol\beta}
\newcommand{\G}{\mathrsfs G}
\renewcommand{\d}{\, \mathrm d}

\makeatother

\usepackage{babel}

\begin{document}

\title{Non-Linear Compton Scattering of Ultrashort and Ultraintense Laser
Pulses}

\author{D. Seipt}

\email{d.seipt@fzd.de}

\author{B. K{\"a}mpfer}

\email{b.kaempfer@fzd.de}

\affiliation{Forschungszentrum Dresden-Rossendorf, POB 51 01 19, 01314 Dresden,
Germany}


\begin{abstract}
The scattering of temporally shaped intense laser pulses off electrons
is discussed by means of manifestly covariant quantum electrodynamics.
We employ a framework based on Volkov states with a time dependent
laser envelope in light-cone coordinates within the Furry picture.
An expression for the cross section is constructed, which is independent
of the considered pulse shape and pulse length. A broad distribution
of scatted photons with a rich pattern of subpeaks like that obtained
in Thomson scattering is found. These broad peaks may overlap at sufficiently
high laser intensity, rendering inappropriate the notion of individual
harmonics. The limit of monochromatic plane waves as well as the classical
limit of Thomson scattering are discussed. As a main result, a scaling
law is presented connecting the Thomson limit with the general result
for arbitrary kinematics. In the overlapping regions of the spectral
density, the classical and quantum calculations give different results,
even in the Thomson limit. Thus, a phase space region is identified
where the differential photon distribution is strongly modified by
quantum effects.
\end{abstract}
\maketitle

\section{Introduction}

The use of chirped-pulse amplification \cite{mourou1985} has led
to a prodigious advance in available laser power. The current records
reach several petawatts, and accompanying interest in strong-field
physics culminates in planned large-scale laser facilities such as
the anticipated {}``Extreme Light Infrastructure'' (ELI) \cite{eli}.
The pioneering theoretical studies in strong-field physics considered
both pair creation in a strong field \cite{Reiss:1962} and the cross
channel process, electron photon scattering
\cite{Nikishov:1963,Nikishov:1964a,Nikishov:1964b,Goldman:1964,Narozhnyi:1964,Brown:1964zz,Kibble:1965zz},
dubbed non-linear Compton scattering, where the use of laser beams
has already been suggested. Since then there has been a wealth of
theoretical papers and we refer the reader to the reviews
\cite{McDonald:1986zz,Fernow,Lau:2003,Mourou:2006zz,Salamin:2006ff,Marklund:2006my}.
In non-linear Compton scattering
\begin{equation}
e(p)+\ell\gamma_{L}(k)\rightarrow e'(p')+\gamma(k')\label{reaction.nlcompton}
\end{equation}
a number of $\ell$ photons with momentum $k$ from a high intensity
laser, scatter off an electron with momentum $p$. A convenient measure
of laser intensity is the dimensionless laser amplitude $a\equiv eE/m\omega$,
with $E$ being the root-mean-square electric field and $\omega$
the laser frequency. The parameter $a$ is a purely classical quantity,
representing the work performed by the field on the electron in one
wavelength. Thus, $a$ is the classical nonlinearity parameter \cite{ritus_doctorskaya}
and it is related to the ponderomotive potential $U_{p}=ma^{2}/2$.
The definition of $a$ can be made explicitly Lorentz and gauge invariant
\cite{Heinzl:2008rh}. When $a$ becomes of order unity the quiver
motion of the electron in the laser beam becomes relativistic in a
classical picture. 

The spectrum of non-linear Compton scattering has been observed in
several experiments colliding laser and electron beams, such as low-intensity
laser photons ($a=0.01$) with low-energy ($\sim1\ {\rm keV}$) electrons
from an electron gun \cite{Englert:1983zz}, $a=2$ photons with plasma
electrons from a gas jet \cite{Umstadter} and, more recently, sub-terawatt
photons ($a=0.35$) from a ${\rm CO_{2}}$ laser with $60$ MeV electrons
from a linac at the BNL-ATF \cite{Babzien:2006zz}. Using linearly
polarized photons the latter two experiments \cite{Umstadter,Babzien:2006zz}
have analyzed the characteristic azimuthal intensity distributions
confirming quadrupole and sextupole patterns for the second and third
harmonics, respectively. Recently, the energy spectrum of the scattered
radiation has been measured in an all-optical setup using laser accelerated
electrons \cite{schwoerer:2006}.

Probably the best known experiment is SLAC E-144 probing strong-field
QED using a terawatt Nd:glass laser ($a\simeq0.6$) in conjunction
with high-energy ($46.6$ GeV) electrons \cite{Bamber:1999zt}. The
observation of non-linear Compton scattering has been reported \cite{Bula:1996st}
as well as the observation of the crossed process
of non-linear pair creation, due to the interaction of a Compton scattered
high-energy photon with a second laser beam \cite{Burke:1997ew}.

The low-energy limit (in terms of laser frequency) of Compton scattering
is Thomson scattering which is described completely classically \cite{Schappert,Esarey}.
This classical picture is used as theoretical framework for many applications
of laser Compton scattering such as X-ray sources \cite{schoenlein,chouffani,debus}
or diagnostic tools \cite{leemans}. A convenient parameter to distinguish
the two regimes is the quantity
\begin{eqnarray}
y_{\ell} & = & \frac{s_{\ell}-m^{2}}{m^{2}}=\frac{2\ell k\cdot p}{m^{2}},\label{eq.def.yl}
\end{eqnarray}
where $s_{\ell}=(p+\ell k)^{2}$ expresses the center of mass energy
squared for the generation of the $\ell$th harmonic in a Lorentz
invariant manner\footnote{$p$ an $k$ are four-vectors, thus $k\cdot p$ denotes a scalar product;
we employ units with $\hbar=c=1$.}.
The $\ell+1\to2$ process is kinematically equivalent to the scattering
of one photon with momentum $\ell k$ off an electron with momentum
$p$, thus it appears as a (pseudo) $2\to2$ process. The Thomson
regime is recovered for $y_{\ell}\ll1$, while for $y_{\ell}>1$ one
finds striking differences to the Thomson scattering. The electron recoil during the scattering
may be quantified by the Lorentz-invariant quantity $t=(p-p')^{2}$ which
is in the range $0\leq-t\le m^{2}\frac{y_{\ell}^{2}}{1+y_{\ell}}$,
i.e.~in the Thomson regime $-t/m^{2}\ll1$ holds.

A quantity measuring non-linear quantum effects,
\begin{eqnarray}
\chi_{R} & = & \frac{e\sqrt{(F^{\mu\nu}p_{\nu})^{2}}}{m^{3}}=\frac{1}{2}ay_{1},
\end{eqnarray}
has been introduced in \cite{Nikishov:1963,Narozhnyi:1964}. It measures
the work done by the field over the Compton wavelength $m^{-1}$ in
the rest frame of the initial electron where the four-vector $p^{\mu}=(m,0,0,0)$.
Introducing the critical field strength \cite{sauter} $E_{S}=m^{2}/e=1.3\times10^{18}\ {\rm V/m},$
this may also be written as $\chi_{R}=E_{\star}/E_{S}$, where $E_{\star}$
is the rms electric field strength in the electron's rest frame. The
parameter $\chi_{R}$ combines nonlinearity and quantum effects. $\chi_{R}$
is of the order of unity if both $a$ and $y_{1}$ are of the order
of unity. Thus, the corrections to the classical description (Thomson
scattering) are important if either ($i$) an ultraintense high-energy
photon pulse, e.g.~produced by an X-ray free electron laser, interacts
with low-energy electrons, or ($ii$) a multi-GeV electron beam is
brought to collision with an optical high-intensity laser. The latter
scenario is similar to the SLAC E-144 experiment but with a higher
value of $a$. For the $50\ {\rm GeV}$ SLAC beam in conjunction with
a counterpropagating optical laser ($\omega\sim1\ {\rm eV}$) one
has $y_{\ell}\sim1$. These parameters will be used mainly below for
numerical calculations. The FACET project \cite{facet} at SLAC envisages
investigations within such kinematics in line with ($ii$).

Ultraintense lasers use short pulses (few fs, few laser cycles) requiring
a proper treatment of the laser pulse structure. Rich substructures
of the scattered photon spectra were predicted within the classical
picture \cite{Gao,Krafft,HartemannPRE,SeiptHeinzl} of Thomson scattering.
These substructures have not yet been confirmed experimentally. The
effect of radiation back reaction on the spectra was studied in \cite{hartemannprl}
and found to be important. Only a few publications address quantum
calculations in pulsed fields for scalar particles \cite{Neville}
and for spinor particles \cite{fofanov,Boca}. In \cite{Piazza},
a connection between the emitted angular spectrum in non-linear Compton
scattering and the carrier envelope phase in few-cycle laser pulses
was established. In a related field, electron wave-packet dynamics in strong
laser fields has been studied in e.g.~\cite{Keitel1,Keitel2}.

In our paper we calculate the emitted photon spectrum in non-linear
Compton scattering using a generalized Volkov solution with temporal
shape in light-cone coordinates. For this purpose we first focus on
the structure of the Volkov wavefunction in a pulsed laser field.
The aim of our study is to compare the QED calculations for non-linear
Compton scattering with results from a classical calculation, i.e.
Thomson scattering.

Our paper is organized as follows. In section \ref{sect.volkov.states}
we present Volkov states in pulsed laser fields. Section \ref{sect.matrix.element}
continues with the calculation of the matrix element and the transition
probability. A slowly varying envelope approximation is discussed.
Numerical results are presented in section \ref{sect.results}. We
discuss various limiting cases of our general results, including monochromatic
plane waves and the Thomson limit. As a main result, a scaling law
is presented, connecting the Thomson spectrum with the Compton spectrum.
In the Appendix we summarize the kinematics in light cone coordinates
and present the Fourier transformation of the Volkov state.

\section{Temporally Shaped Volkov States}

\label{sect.volkov.states}

A strong laser field may be considered as a coherent state of photons
$|\mathcal{C}\rangle$, characterized by the polarization and momentum
distribution $\mathcal{C}{}^{\mu}(k)$, if the depletion of the laser
photons from $|\mathcal{C}\rangle$ by an interaction process with
electrons is negligible, i.e.~for any relevant scattering process
$S=\langle{\it out};\mathcal{C}'|S|{\it in};\mathcal{C}\rangle$ with
$\mathcal{C}'=\mathcal{C}$ is valid, where \emph{in} and \emph{out}
are particle number states without coherent parts \cite{HarveyHeinzl}.
Then, it is possible to work within the Furry picture \cite{Furry},
where the interaction of an electron with the classical background
field $A^{\mu}(x)$, which is the Fourier transform of $\mathcal{C}{}^{\mu}(k)$,
is treated nonperturbatively and solutions of the Dirac equation
\begin{eqnarray}
(i\slashed\partial-e\slashed A-m)\psi(x) & = & 0\label{eq.dirac}
\end{eqnarray}
are utilized as basic \emph{in} and \emph{out} states for the
perturbative expansion of the $S$ matrix. For background fields in
the form of plane waves, closed solutions of \eqref{eq.dirac} can
be found,
\begin{align}
\psi_{p,s}(x)=\left(1+\frac{e}{2k\cdot p}\slashed k\slashed A\right)\exp\{iS_{p}(x)\}\frac{u_{p,s}}{\sqrt{2E_{p}}},
\label{eq:volkov.state}
\end{align}
where the free Dirac spinor for momentum $p$ and spin $s$ fulfills
$(\slashed p-m)u_{p,s}=0$ and is normalized to $\bar{u}_{p,s'}u_{p,s}=2m\delta_{ss'}$.
The phase is the classical Hamilton Jacobi action
\begin{align}
S_{p}(x) & =-p\cdot x+f(k\cdot x),
\end{align}
with $f=f_{1}+f_{2}$, $f_{1}=-\int\limits _{\phi_{0}}^{k\cdot x}\d\phi\frac{eA\cdot p}{k\cdot p}$
and $f_{2}=\int\limits _{\phi_{0}}^{k\cdot x}\d\phi\frac{e^{2}A^{2}}{2k\cdot p}$.
Equation \eqref{eq:volkov.state} represents the famous Volkov states,
whose perturbative expansion in terms of interactions with the laser
field is depicted in figure 1 of \cite{HarveyHeinzl} (The expansion
parameter, i.e.~the coupling strength at the vertices, is $a_{0}$
defined below).

For the vector potential we use a real transverse plane wave
\begin{align}
A^{\mu}=A_{0}\, g(k\cdot x)\,(\epsilon_{1}^{\mu}\cos\xi\cos k\cdot x+\epsilon_{2}^{\mu}\sin\xi\sin k\cdot x),
\label{eq:vector.potential}
\end{align}
modified by an envelope function $g$ and fulfilling $A\cdot k=0$
and $k\cdot k=0$. The parameter $\xi$ determines the polarization
of the laser: It is linearly $x\,(y)$ polarized for $\xi=0\,(\xi=\pi/2)$
and circularly polarized for $\xi=\pm\pi/4$. For other values of
$\xi$, the laser is elliptically polarized \cite{Panek}. The vector
potential is normalized such that the mean energy density or the energy
flux $\langle\mathbf{E}^{2}\rangle\propto-A_{\mu}A^{\mu}=g^{2}A_{0}^{2}/2$,
where $\langle\ldots\rangle$ means averaging over the fast oscillations
of the carrier wave, is independent of $\xi$, but the dimensionless
laser amplitude $a$, as defined in the introduction, is time dependent.
A time independent laser strength parameter may be defined by the
normalized peak value of the vector potential $a_{0}=eA_{0}/m$ (with
this definition, $a^{2}=a_{0}^{2}/2$ for $g\equiv1$). The vector
potential \eqref{eq:vector.potential} can also be cast into a complex
form
\begingroup
  \renewcommand*\theequation{$7'$}
  \begin{align}
    A^{\mu}=\frac{A_{0}g(k\cdot x)}{2}(\epsilon_{+}^{\mu}e^{-ik\cdot x}+\epsilon_{-}^{\mu}e^{+ik\cdot x})=\frac{A_{0}g}{2}B^{\mu},
  \end{align}
\endgroup
\setcounter{equation}{7}
with the complex polarization vectors $\epsilon_{\pm}^{\mu}=\cos\xi\epsilon_{1}^{\mu}\pm i\sin\xi\epsilon_{2}^{\mu}$
with $\epsilon_{+}\cdot\epsilon_{-}=-1$, $\epsilon_{\pm}\cdot\epsilon_{\pm}=\sin^{2}\xi-\cos{}^{2}\xi$
and the definition $B^{\mu}=\epsilon_{+}^{\mu}e^{-ik\cdot x}+\epsilon_{-}^{\mu}e^{+ik\cdot x}$.

In what follows, the temporal pulse shape will often be chosen as
a \emph{cosh} pulse
\begin{align}
 g(k\cdot x) & =   \frac{1}{\cosh
	      \left(
		   \displaystyle{ \frac{k\cdot x}{\sigma} }
	      \right)}, \label{eq:def.sech}
\end{align}
with width $\sigma$, or a Gaussian\begin{equation}
g(k\cdot x)=\exp\left\{ -\frac{(k\cdot x)^{2}}{2\sigma^{2}}\right\} .\label{eq:def.gauss}\end{equation}
Besides these special cases, any other smooth function $g$ which
depends solely on $k\cdot x$ is possible.

\begin{figure}[t]
\begin{centering}
\includegraphics[height=6cm]{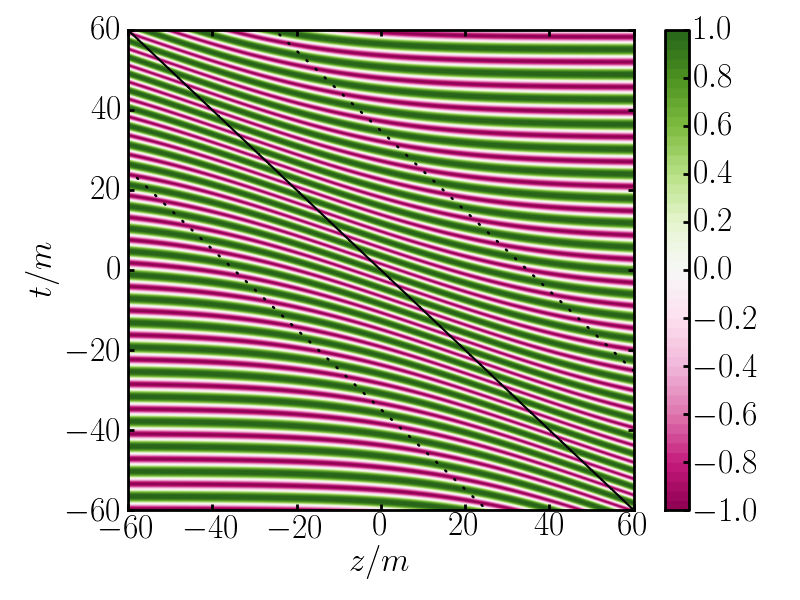}
\includegraphics[height=6cm]{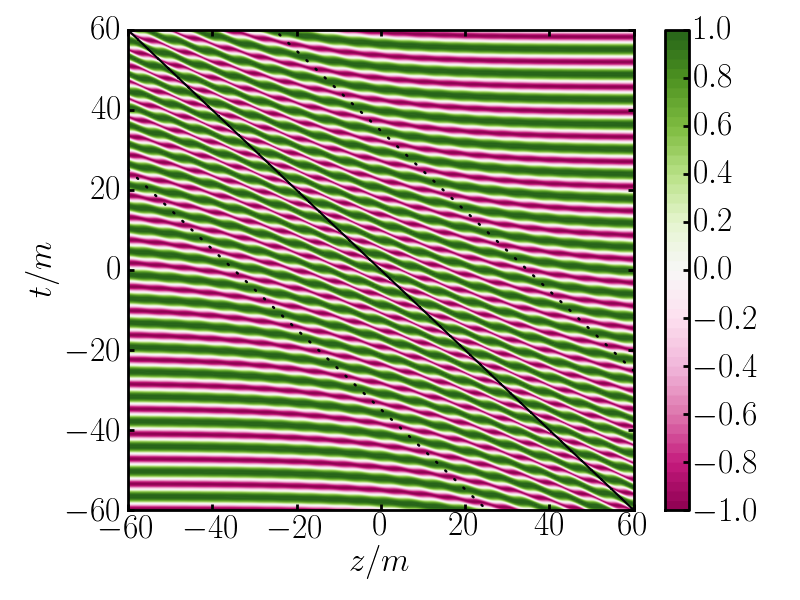}
\par\end{centering}

\begin{centering}
\includegraphics[height=6cm]{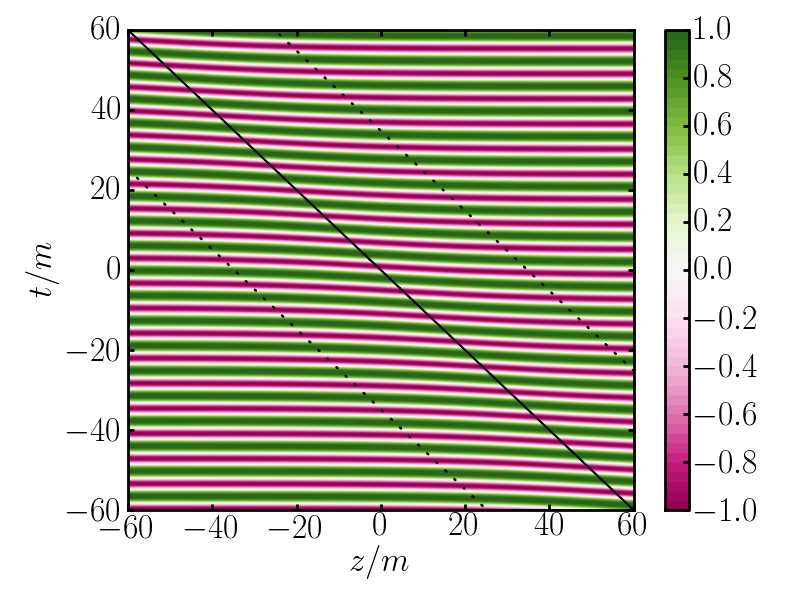}
\includegraphics[height=6cm]{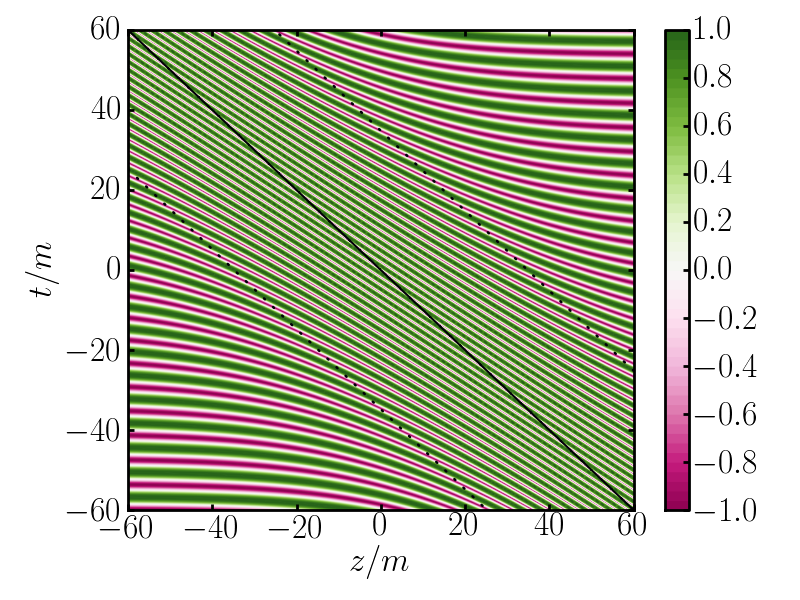}
\par\end{centering}

\caption{Contour plot of $\frac{1}{4}{\rm tr}\, C_{p}(x)$ in position space
in the $z-t$ plane. The laser pulse with \textit{cosh} profile is
located between the two dotted lines. Left (right) top panel: Circularly
(linearly) polarized laser pulse with $a_{0}=1.5$ and $\sigma=20$,
Bottom panels: circular polarization for $a_{0}=0.5$ (left) and $a_{0}=3.0$
(right).}

\label{fig.volkov.state} 
\end{figure}

Since the vector potential $A^{\mu}$ depends only on $k\cdot x$,
it is convenient to work in light-cone coordinates with $k\cdot x=\omega x_{+}$
(see Appendix A). In these coordinates, the Volkov wavefunction \eqref{eq:volkov.state}
reads
\begin{align}
\psi_{p,s}(x)                         & =C_{p}(\mathbf{x}_{\perp},x_{-},x_{+})\frac{u_{p,s}}{\sqrt{2E_{p}}},\\
C_{p}(\mathbf{x}_{\perp},x_{-},x_{+}) & = 
			\Big[ 1+d_{p}g(x_{+})
		    \left(
			\slashed k\slashed\epsilon_{-}e^{i\omega x_{+}}+\slashed k\slashed\epsilon_{+}e^{-i\omega x_{+}}
		    \right)
			\Big]
		    e^{-\frac{i}{2}\left(p_{+}x_{-}+p_{-}x_{+}\right)+i\mathbf{p}_{\perp}\cdot\mathbf{x}_{\perp}+if(x_{+})}
\label{eq.def.B_p}
\end{align}
with $d_{p}=a_{0}m/(4k\cdot p)$. For many purposes it is sufficient
to consider only the function $C_{p}$ since it contains the relevant
information on the interaction of the electron with the laser pulse.
The real part%
\footnote{The imaginary part gives no further information; it has a shifted
phase as compared to the real part. %
} of the scalar projection $\mathcal{S=}\mathrm{\frac{1}{4}tr}\, C_{p}(x)=\exp\{iS_{p}(x)\}$
is visualized in Fig.~\ref{fig.volkov.state} in the frame where the
electron is initially at rest. The scalar projection is essentially
equivalent to probing the state $\psi_{p,s}$ with $\bar{u}_{p,s}$
and an average over the spins $s$
\begin{equation}
\frac{1}{2}\sum_{s}\frac{\bar{u}_{p,s}}{\sqrt{2E_{p}}}\psi_{p,s}(x)=\frac{m}{E_{p}}\mathcal{S}.
\end{equation}
\begin{figure}[t]
\begin{centering}
\includegraphics[height=6cm]{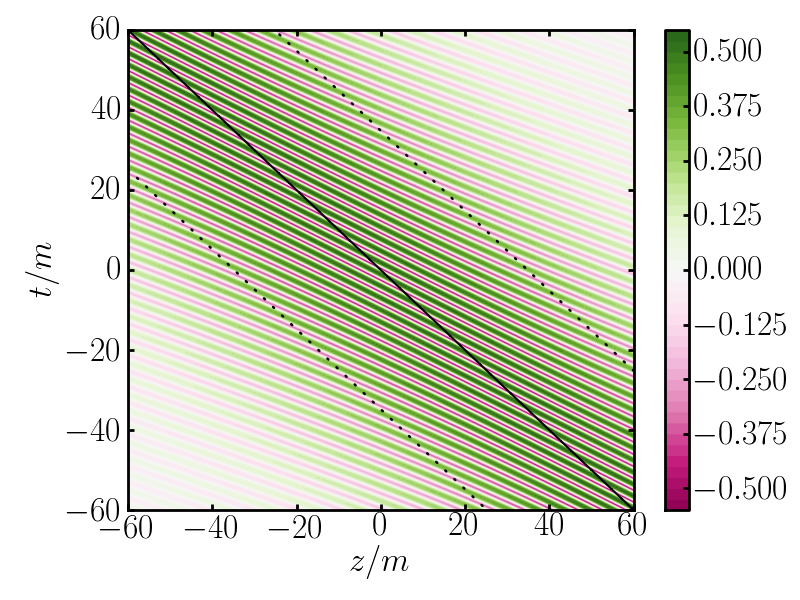}
\includegraphics[height=6cm]{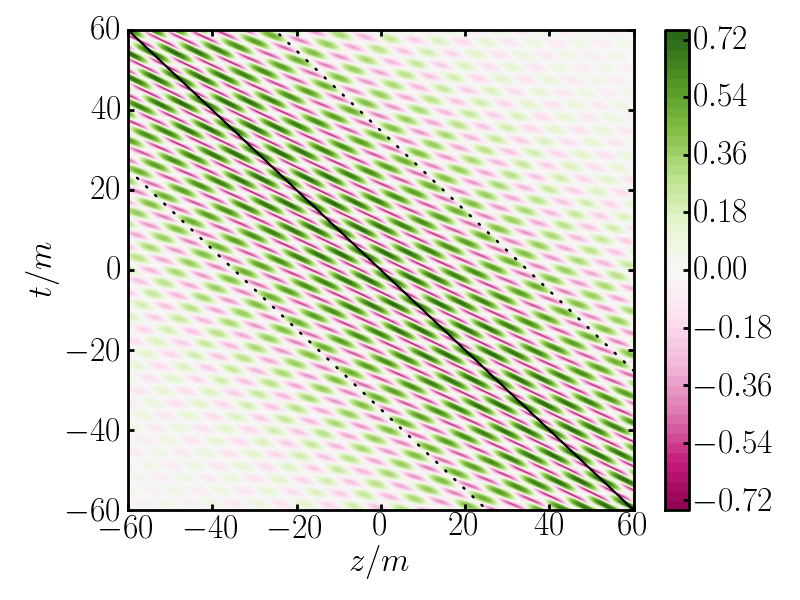}
\par\end{centering}

\begin{centering}
\includegraphics[height=6cm]{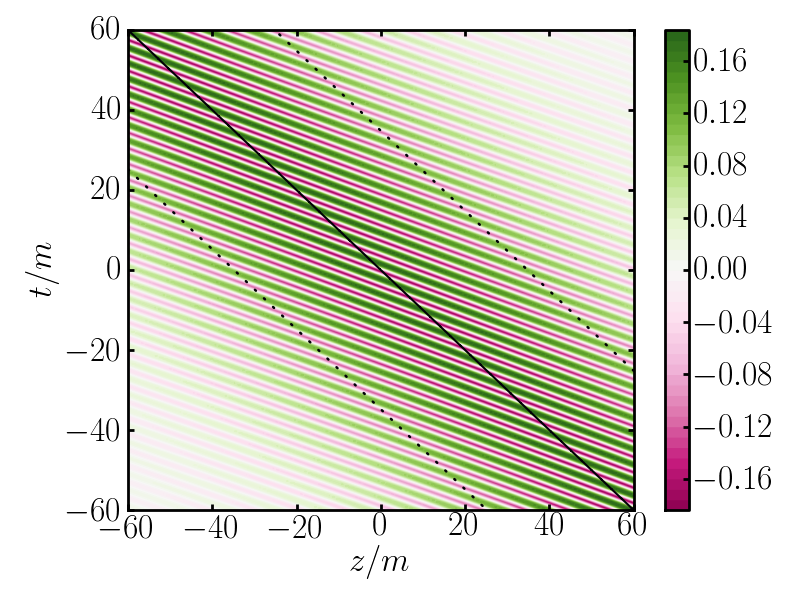}
\includegraphics[height=6cm]{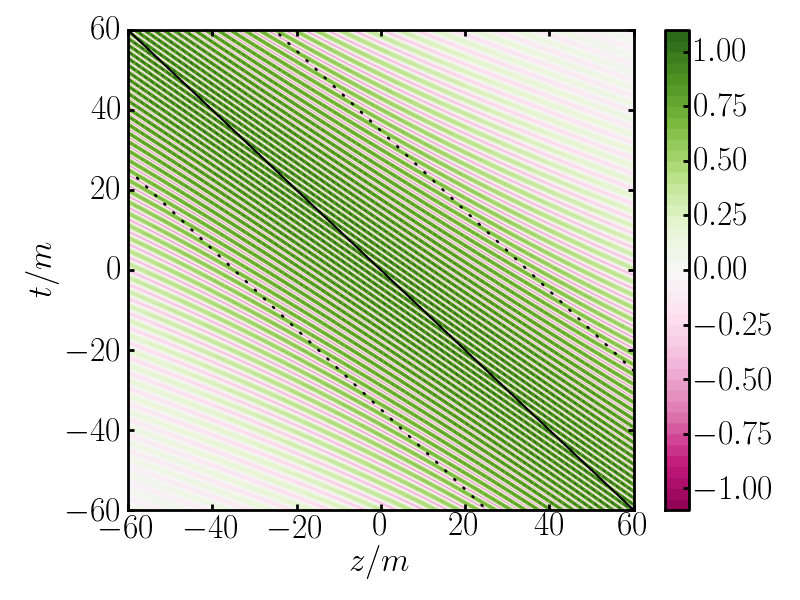}
\par\end{centering}

\caption{Contour plot of the spin flip contribution to the Volkov wave function
by the tensor projection $\mathcal{T}^{13}-i\mathcal{T}^{23}$. Same
parameters as in Fig.~\ref{fig.volkov.state}.}

\label{fig.volkov.state.spinflip} 
\end{figure}
In that frame the (free) electron wavefunction outside the laser pulse
behaves as $\propto e^{-ip\cdot x}=e^{-imt}$. The effect of the laser
pulse is a local deformation of the electron wavefronts due to the
build-up of an effective, time dependent momentum $q^{\mu}(x_{+})=p^{\mu}+g^{2}(x_{+})b_{p}k^{\mu}$
with $b_{p}=m^{2}a_{0}^{2}/4k\cdot p$. The momentum $q^{\mu}(x_{+})$
achieves its maximum at the center of the laser pulse $x_{+}=0$,
depicted as diagonal straight line in Fig.~\ref{fig.volkov.state},
where it coincides with the usual quasi momentum in monochromatic
plane waves. Thus, inside the laser pulse, especially for $x_{+}=0$,
the fully dressed electron wavefunction behaves as $\propto e^{-iq\cdot x}=e^{-i(m+b_{p}\omega_{\star})t+ib_{p}\omega_{\star}z}$,
i.e.~the electron wavelength changes and the wavefronts become tilted.
Both effects are proportional to the ponderomotive potential, i.e.
$\propto b_{p}\omega_{\star}=ma_{0}^{2}/4=U_{p}/2$ , where $\omega_{\star}=u\cdot k$
is the laser frequency in the initial electron rest frame and $u=p/m$.
In Fig.~\ref{fig.volkov.state}, $\omega_{\star}=300\ {\rm keV}$
is chosen which can be achieved, for instance, with laser photons
of $1.5\ {\rm eV}$ energy colliding head-on with $50\ {\rm GeV}$
electrons, or a $15\ {\rm keV}$ X-ray laser beam with $5\ {\rm MeV}$
electrons. The electron wavefunction changes its behavior from the
free case to the fully dressed case over $N =\sigma m / (2\pi \omega_\star)$
oscillations of the free electron wavefunction, i.e.~$N=5.5$ for the parameters
employed in Fig.~\ref{fig.volkov.state}. For lower values of $\omega_\star$,
the behavior of the electron wavefunction changes slowly over many oscillations,
e.g.~for $\omega_\star = 200 \ \rm eV$ ($35\ \rm MeV$ electrons colliding with $1.5 \ \rm eV$ photons)
and $\sigma=20$ one finds $N=8100$.
For a linearly polarized
laser, additional ripples appear in the interaction region due to
the oscillating terms in the phase. For non-head-on geometries, similar
ripples are also present for circular polarization. The smoothness
of the pulse envelope $g$ ensures the smoothness of the Volkov wavefunction
in the transition from the field-free regions to the laser pulse.

Vector projections with an odd number of Dirac matrices, such as $\mathcal{V^{\mu}=}\frac{1}{4}\mathrm{tr}\,\gamma^{\mu}C_{p}(x)$
or $\mathcal{\mathcal{A}^{\mu}=}\frac{1}{4}\mathrm{tr}\,\gamma^{5}\gamma^{\mu}C_{p}(x)$,
vanish identically and are, therefore, not useful for characterizing
$\psi_{p,s}$. The pseudoscalar $\mathcal{P}=\frac{1}{4}\mathrm{tr}\,\gamma^{5}C_{p}(x)$
also vanishes. The antisymmetric tensor projections $\mathcal{T}^{\mu\nu}=\frac{1}{4}\mathrm{tr}\,\sigma^{\mu\nu}C_{p}(x)=id_{p}(k^{\nu}A^{\mu}-k^{\mu}A^{\nu})\exp\{iS_{p}(x)\}$,
where $\sigma^{\mu\nu}=\frac{i}{2}[\gamma^{\mu},\gamma^{\nu}]$ is
the spin tensor, allow for a further characterization by $\sigma^{01},\sigma^{02},\sigma^{13},\sigma^{23}$
for circular laser polarization and for $\sigma^{01},\sigma^{13}$
for linear polarization. These tensor projections are nonzero but only inside the laser pulse. They
mix contributions with different spin orientation and
are therefore proportional to a combination of $\bar{u}_{p,1}\psi_{p,2}(x)$
and $\bar{u}_{p,2}\psi_{p,1}(x)$, i.e.~the spin-up wavefunction contains
contributions with spin-down and vice versa. From the structure of
the Pauli interaction term $\sigma^{\mu\nu}F_{\mu\nu}$, where $F_{\mu\nu}=\partial_{\mu}A_{\nu}-\partial_{\nu}A_{\mu}$
is the electromagnetic field strength tensor, one can infer that $\sigma^{01}\,(\sigma^{02})$
corresponds to the interaction of the electron with the $x\,(y)$
component of the electric field, and $\sigma^{13}\,(\sigma^{23})$
corresponds to the $y\,(x)$ component of the magnetic field. From
this correspondence and by inspecting \eqref{eq:vector.potential}
it is easy to understand why some projections are zero for linear
polarization while they are nonzero for circular polarization. As
an example, the tensor projection $\mathcal{T}^{13}-i\mathcal{T}^{23}$
is shown in Fig.~\ref{fig.volkov.state.spinflip}.

\section{Calculation of the Matrix Element}

\label{sect.matrix.element}

\subsection{The $S$ matrix}

\begin{figure}[t]
\noindent \begin{centering}
\includegraphics[width=6cm]{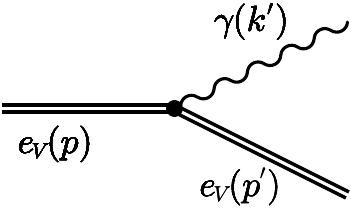}
\par\end{centering}

\caption{Feynman diagram for non-linear Compton scattering as the decay of
a laser dressed Volkov electron state. }
\label{fig.feynman}
\end{figure}

The interaction of the Volkov electron $e_{V}$ with photon modes
different from the laser field are treated by perturbative $S$ matrix
expansion. The Born approximation of the matrix element for the emission
of one photon, i.e.~non-linear Compton scattering $e_{V}(p)\to e_{V}(p')+\gamma(k')$,
is depicted in Fig.~\ref{fig.feynman}. Using Feynman rules \cite{Landau4},
the $S$ matrix element for such a process is given by
\begin{align}
S_{fi}=\langle p',s';k',\epsilon_{\lambda'}'|S[A]|p,s\rangle & =
	  -ie\int\d^{4}x\bar{\psi}_{p',s'}(x)\frac{e^{ik'\cdot x}}{\sqrt{2\omega'}}\slashed\epsilon'_{\lambda'}\psi_{p,s}(x),
\label{eq:matrix.element}
\end{align}
which reads in light-cone coordinates, suppressing spin $(s,s')$
and polarization $(\lambda')$ indices from now on,
\begin{eqnarray}
S_{fi} & = & N_{0}\int\d^{4}x\bar{u}_{p'}\big(1+d_{p'}\slashed B\slashed k\big)\slashed\epsilon'
		\big(1+d_{p}\slashed k\slashed B\big)u_{p}e^{i(S_{p}-S_{p'}+ik'\cdot x)} \\
       & = & \frac{N_{0}}{2}\int\d^{2}\mathbf{x}_{\perp}\d x_{+}\d x_{-}\,\Gamma(x_{+})e^{iH(x_{+},x_{-},\mathbf{x_{\perp}})}
\label{eq.matrixelemet.lightcone}
\end{eqnarray}
with $N_{0}=-ie/\sqrt{2\omega'2E_{p}2E_{p'}}$ and
\begin{eqnarray}
\Gamma(x_{+}) & = & \mathrsfs T_{0}^{0}
		    + g\, e^{i\omega x_{+}}\mathrsfs T_{1}^{1}+g\, e^{-i\omega x_{+}}\mathrsfs T_{-1}^{1}
		    + g^{2}\,\mathrsfs T_{0}^{2}
                    +g^{2}\, e^{2i\omega x_{+}}\mathrsfs T_{2}^{2}+g^{2}\,e^{-2i\omega x_{+}} \mathrsfs T_{-2}^{2},
\end{eqnarray}
where
\begin{align}
\mathrsfs T_{0}^{0} & =\bar{u}_{p'}\slashed\epsilon'u_{p},\\
\mathrsfs T_{\pm1}^{1} & =\bar{u}_{p'}\left(d_{p'}\slashed\epsilon_{\mp}\slashed k\slashed\epsilon'+d_{p}\slashed\epsilon'\slashed k\slashed\epsilon_{\mp}\right)u_{p},\\
\mathrsfs T_{0}^{2} & =4(k\cdot\epsilon')d_{p}d_{p'}\bar{u}_{p'}\slashed ku_{p},\\
\mathrsfs T_{\pm2}^{2} & =d_{p}d_{p'}\bar{u}_{p'}\left({\slashed\epsilon_{\mp}\slashed k\slashed\epsilon'\slashed k\slashed\epsilon_{\mp}}\right)u_{p}.
\end{align}
Due to $\slashed\epsilon_{\mp}\slashed k\slashed\epsilon'\slashed k\slashed\epsilon_{\mp}=-2(\epsilon'\cdot k)(\epsilon_{\mp}\cdot\epsilon_{\mp})\slashed k$,
one finds $\mathrsfs T_{\pm2}^{2}=0$ for circular polarization. Furthermore,
\begin{eqnarray}
H(x_{+},x_{-},\mathbf{x_{\perp}}) & = & S_{p}-S_{p'}+ik'\cdot x\nonumber \\
 & = & (p'+k'-p)\cdot x+f(x_{+})-f'(x_{+})\label{eq.def.H}
\end{eqnarray}
with $f'=f(p\to p')=f_{1}(p')+f_{2}(p')$ and
\begin{align}
f_{1}(x_{+};p) & =\frac{ma_{0}}{k\cdot p}\int\limits _{\phi_{0}}^{k\cdot x}\d\phi\, g(\phi)\left[p\cdot\epsilon_{1}\cos\xi\cos\phi+p\cdot\epsilon_{2}\sin\xi\sin\phi\right],\label{eq.f1}\\
f_{2}(x_{+};p) & =-\frac{m^{2}a_{0}^{2}}{2k\cdot p}\int\limits _{\phi_{0}}^{k\cdot x}\d\phi\, g^{2}(\phi)\big[\cos^{2}\xi\cos^{2}\phi+\sin^{2}\xi\sin^{2}\phi\big].\label{eq.f2}
\end{align}
Inspecting Eq.~\eqref{eq.def.H}, it is obvious that the dependence
of $H$ on $x_{-}$ and $\mathbf{x}_{\perp}$ is trivial and the integrations
over these variables in Eq.~\eqref{eq.matrixelemet.lightcone} can
be done analytically. As a result, energy-momentum conservation is
imposed on the components $P_{+}$ and $\mathbf{P}_{\perp}$, and
the exponent
\begin{eqnarray}
H_{+}(x_{+}) & = & \frac{1}{2}(k'_{-}+p'_{-}-p_{-})x_{+}+f(x_{+})-f'(x_{+})\label{eq.def.Hplus}
\end{eqnarray}
remains. Due to the non-trivial pulse dependent structure of $H_{+}$,
the $x_{+}$ integration does not yield another conservation law.
Thus, the frequency of scattered photons $\omega'$ is not fixed by
energy and momentum conservation as a function of scattering angle
$\theta$ and remains as independent parameter. Including the $x_{+}$
dependence of $\Gamma(x_{+})$, some rather complicated functions
of $\omega'$ emerge
\begin{align}
\mathrsfs A_{N}^{M} & ={\displaystyle \int\limits _{-\infty}^{\infty}\d x_{+}g^{M}(x_{+})\exp i\{H_{+}(x_{+})+N\omega x_{+}\}.}
\label{eq.def.A_functions}
\end{align}
With these definitions, the $S$ matrix element can be written as
\begin{align}
S_{fi} & =(2\pi)^{3}\delta^{2}({\bf k'_{\perp}+p'_{\perp}-p_{\perp}})\delta(k'_{+}+p'_{+}-p_{+})N_{0}\mathrsfs M
\label{eq.S_matrix.final}
\end{align}
with
\begin{align}
\mathrsfs M & =\mathrsfs T_{0}^{0}\mathrsfs A_{0}^{0}+\mathrsfs T_{1}^{1}\mathrsfs A_{1}^{1}+\mathrsfs T_{-1}^{1}\mathrsfs A_{-1}^{1}+\mathrsfs T_{0}^{2}\mathrsfs A_{0}^{2}+\mathrsfs T_{2}^{2}\mathrsfs A_{2}^{2}+\mathrsfs T_{-2}^{2}\mathrsfs A_{-2}^{2}.
\label{eq.def.M_matrix}
\end{align}
The integrals $\mathrsfs A_{N}^{M}$ are numerically convergent for
$M\geq1$ due to the presence of the pulse function in the integrand,
rendering the range of integration practically finite. The integral
$\mathrsfs A_{0}^{0}$, however, contains a divergent part and must
be regularized. A possible method has been proposed in \cite{Boca},
where one multiplies the integrand with a convergence factor $e^{-\varepsilon|x_{+}|},\:\varepsilon>0$,
and performs an integration by parts. The result is
\begin{eqnarray}
\mathrsfs A_{0}^{0} & = & -\frac{2}{P_{-}}\intop_{-\infty}^{\infty}\d x_{+}\frac{\d(f-f')}{\d x_{+}}\exp\{iH_{+}(x_{+})\}+4e^{i[f(0)-f'(0)]}\lim_{\varepsilon\to0^{+}}\frac{\varepsilon}{P_{-}^{2}+\varepsilon^{2}},
\label{eq.a00.reg}
\end{eqnarray}
with $P_{-}=k'_{-}+p'_{-}-p_{-}$. In \eqref{eq.a00.reg}, the first
part is now convergent and the second part is proportional to a $\delta$
distribution with support at $\omega'=0$. The latter contribution
can be neglected in our analysis for $\omega'>0$. This regularized
version of $\mathrsfs A_{0}^{0}$ will be used in the subsequent numeric
calculations.

\subsection{Slowly Varying Envelope Approximation}

The calculations are simplified upon utilizing the slowly varying
envelope approximation (SVEA) of the phase of the $\mathrsfs A_{N}^{M}$
functions. This approximation scheme is suitable for long pulses with
$\sigma\gg1$. Typically $\sigma$ is proportional to the number of
laser oscillations under the envelope. For $f_{1}$, which is proportional
to $g$ (see Eq.~\eqref{eq.f1}), an integration by parts is performed,
yielding
\begin{eqnarray}
\int\d\phi g(\phi)\sin\phi & = & -g(\phi)\cos\phi+\int\d\phi\frac{\d g}{\d\phi}\cos\phi,\\
\int\d\phi g(\phi)\cos\phi & = & g(\phi)\sin\phi-\int\d\phi\frac{\d g}{\d\phi}\sin\phi.
\end{eqnarray}
SVEA basically means neglecting the second terms containing the derivative
of the pulse shape $\d g/\d\phi$ because it is $\mathcal{O}(1/\sigma)$
smaller than the first term. For $f_{2}$ ($\propto g^{2}$, cf.~Eq.~\eqref{eq.f2}) we use
\begin{eqnarray}
\int\d\phi g^{2}(\phi)\cos^{2}\phi & \approx & \frac{1}{2}\int\d\phi g^{2}(\phi)+\frac{1}{2}g^{2}(\phi)\sin\phi\cos\phi,\\
\int\d\phi g^{2}(\phi)\sin^{2}\phi & \approx & \frac{1}{2}\int\d\phi g^{2}(\phi)-\frac{1}{2}g^{2}(\phi)\sin\phi\cos\phi,
\end{eqnarray}
which becomes particularly handy if $\int\d\phi\, g^{2}$ is known
analytically, such as for the \emph{sech} pulse \eqref{eq:def.sech},
where $\int\d\phi\cosh^{-2}\phi/\sigma=\sigma\tanh(\phi/\sigma)+const$,
or the Gaussian pulse \eqref{eq:def.gauss}, $\int\d\phi\exp(-\phi^{2}/2\sigma^{2})^{2}=\sqrt{\pi}\sigma{\rm erf}(\phi/\sigma)/2+const$,
where ${\rm erf}(x)$ is the normalized error function. Finally, the
SVEA result for the phase reads
\begin{eqnarray}
f_{1} & = & \frac{ma_{0}}{k\cdot p}g(x_{+})\left[p\cdot\epsilon_{1}\cos\xi\sin\omega x_{+}-p\cdot\epsilon_{2}\sin\xi\cos\omega x_{+}\right],\\
f_{2} & = & -\frac{m^{2}a_{0}^{2}}{4k\cdot p}\Bigg[\intop^{\omega x_{+}}\d\phi g^{2}(\phi)+g^{2}(x_{+})\cos\omega x_{+}\sin\omega x_{+}(\cos^{2}\xi-\sin^{2}\xi)\Bigg],
\end{eqnarray}
generalizing the approximation scheme of \cite{fofanov} to linear
laser polarization.

\begin{figure}[t]
\noindent \begin{centering}
\includegraphics[angle=270,width=16cm]{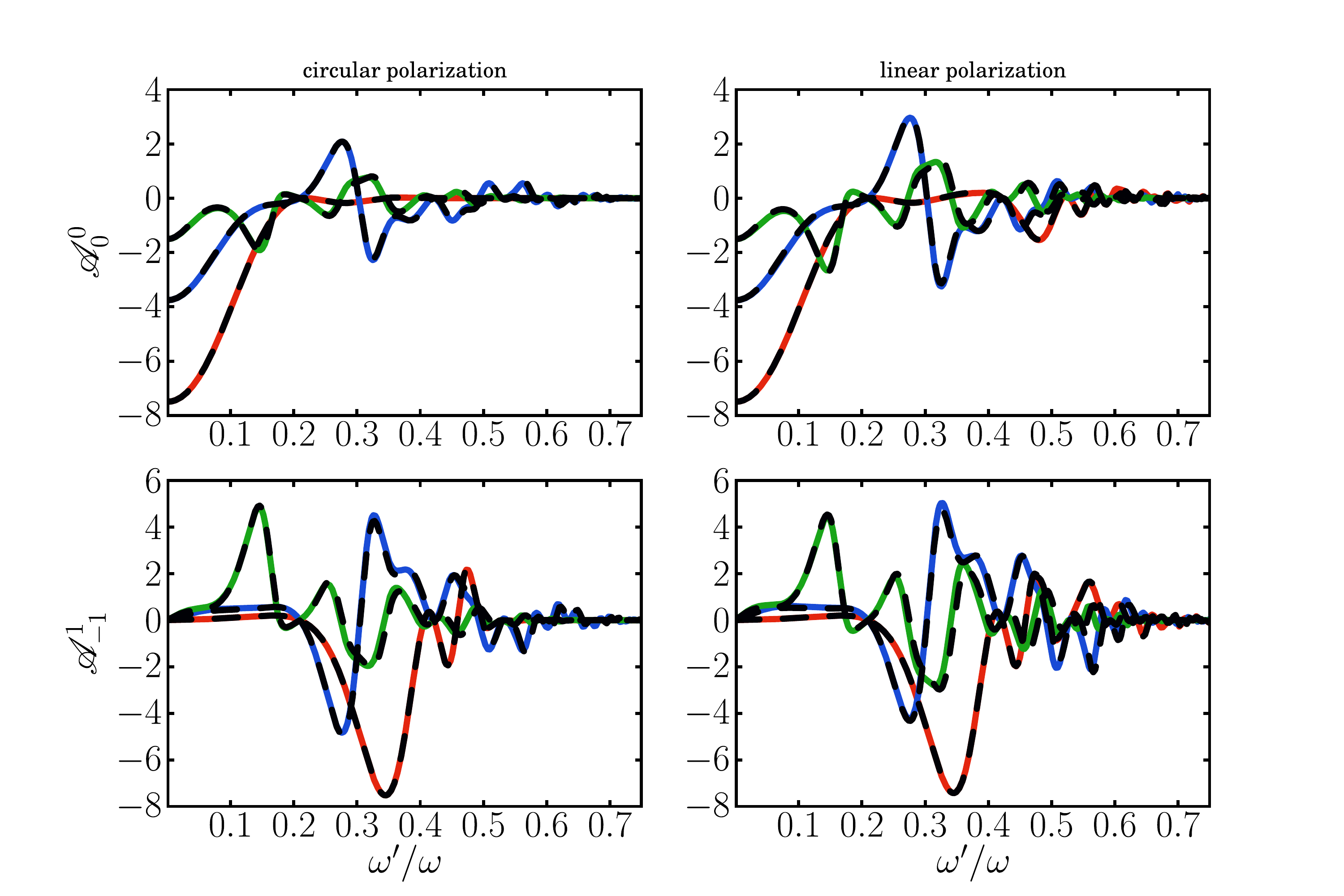}
\par\end{centering}

\caption{Comparison of the SVEA (black dashed curves) and the full numerical
results (solid curves for $\theta=0$ (red), $\theta=1/\gamma$ (blue)
and $\theta=2/\gamma$ (green)) for the real parts of the functions
$\mathrsfs A_{0}^{0}$ and $\mathrsfs A_{-1}^{1}$ for $\sigma=5$.
Parameters are $a_{0}=1.5$, $\omega=1.5\ {\rm eV}$, $\gamma=10^{5}$.
Left (right) panels are for circular (linear) polarization.}
\label{fig.SVEA.validity}
\end{figure}

Even for short pulses, such as for $\sigma=5$ meaning that there
are about $5$ laser oscillations in the pulse, i.e.~the pulse length
is $\approx15\,{\rm fs}$ for $\lambda=800\,{\rm nm}$, SVEA is quite
a good approximation, see Fig.~\ref{fig.SVEA.validity} for selected
examples.

\subsection{The spectral distribution of scattered photons and the cross section}

\label{sect.cross.section}

\begin{figure}[t]
\noindent \begin{centering}
\includegraphics[angle=270,width=10cm]{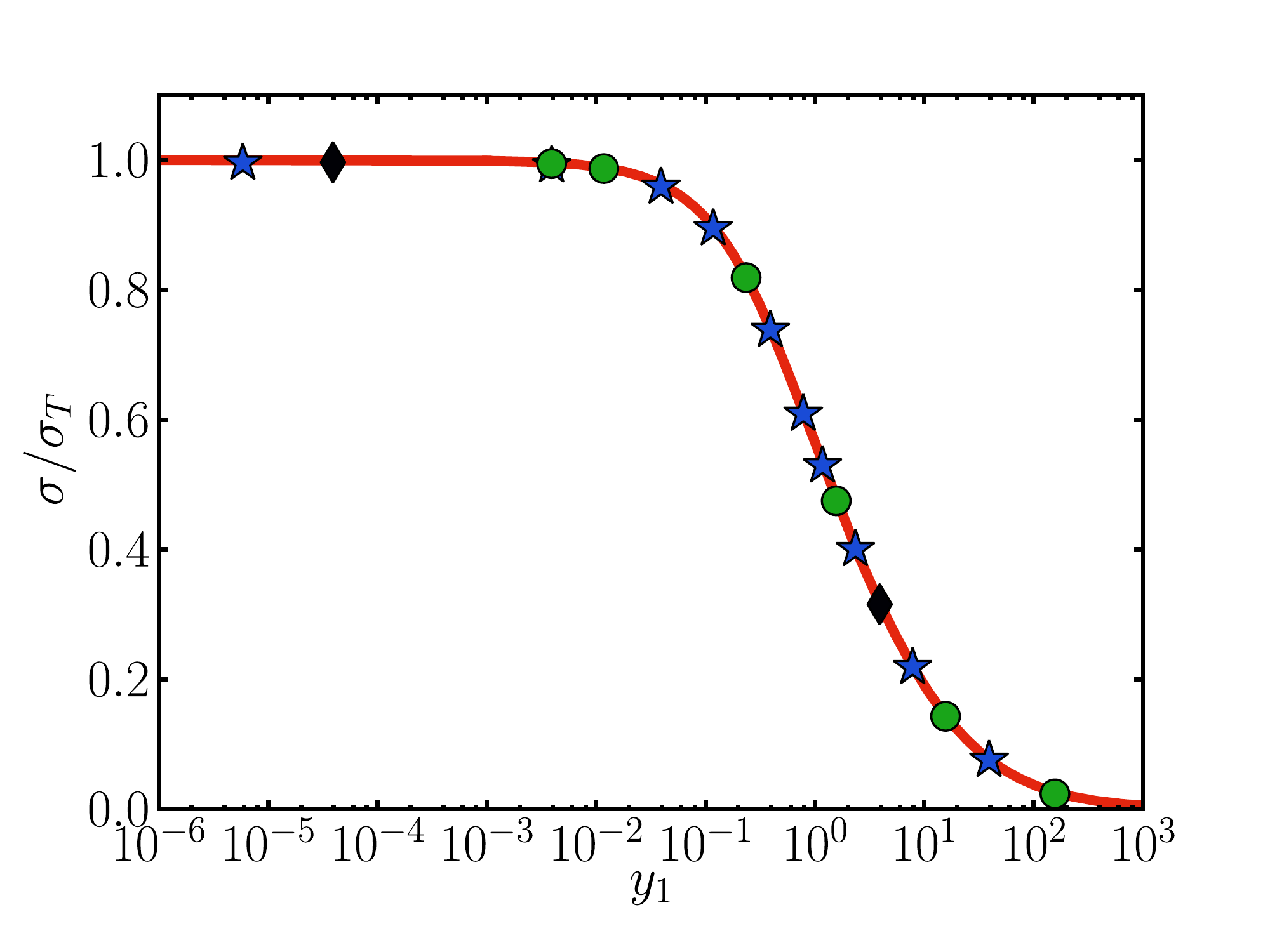}
\par\end{centering}

\caption{Total cross section for Compton scattering normalized to the Thomson
cross section $\sigma_{T}$. Red curve: Klein-Nishina cross section.
Symbols: numerically calculated cross section in a pulsed laser field
with $a_{0}=0.001$. Blue stars: $\sigma=20$ for a \emph{cosh}
pulse; green circles: $\sigma=30$ for a Gaussian pulse; black diamonds:
$\sigma=100$ for a Gaussian pulse. }

\label{fig.towards.plane.wave}
\end{figure}

In the standard formalism, scattering experiments are thought of as
constant streams of particles interacting. Consequently, the square
of the $S$ matrix contains a factor $T$ which originates from the
square of the energy-momentum conservation which is interpreted as
$\delta(P_{i}-P_{f})^{2}\to\frac{VT}{(2\pi)^{4}}\delta(P_{i}-P_{f})$,
with the volume $V$ and interaction time $T$ which are both put
to infinity. On the purpose of rendering this quantity finite, usually
the differential rate per unit time and unit volume $\d w_{i\to f}=\frac{|S_{fi}|^{2}}{VT}\d\Pi$
is considered, where $\d\Pi$ denotes the final state phase space.
Here, however, the interaction is happening only within a finite time
interval. Because of lacking one $\delta$ distribution, the square
of the $S$ matrix now reads
\begin{equation}
|S_{fi}|^{2}=(2\pi)^{3}V\delta(\mathbf{p}_{\perp}'+\mathbf{k}_{\perp}'-\mathbf{p}_{\perp})\delta(p'_{+}+k'_{+}-p_{+})|N_{0}\mathrsfs M|^{2},
\end{equation}
where the dependence on the interaction time is contained in $\mathrsfs M$
and is finite. Thus, it is not necessary to define a differential
rate per unit time. An appropriate observable is the emission probability
of photons per unit volume and laser pulse
\begin{equation}
\d N=\frac{|S_{fi}|^{2}}{V}\d\Pi,
\end{equation}
which has as classical analog the spectral density of scattered photons
in Thomson scattering (cf.~e.g.~\cite{SeiptHeinzl,Schappert})
\begin{align}
\frac{\d^{2}N_{{\rm classical}}}{\d\omega'\d\Omega} & =-\frac{\omega'}{16\pi^{3}}j^{*}(k')\cdot j(k'),\label{eq.classical.spectral.density}\\
j^{\mu}(k') & =e\int\d\tau u^{\mu}(\tau)e^{ik'\cdot x(\tau)},\label{eq.retarded.current}
\end{align}
where $u^{\mu}(\tau),x^{\mu}(\tau)$ are the classical velocity
and orbit from a solution of the Lorentz force equation for a spinless
pointlike charge, and $j^{\mu}(k')$ is the retarded Fourier transform
of the electron current. The notion of Thomson scattering is specified
to mean this particular calculation scheme. A quantum spectral density
is given by the Lorentz invariant expression
\begin{align}
\frac{\d^{2}N_{ss'\lambda'}}{\d\omega'\d\Omega} & =\frac{e^{2}\omega'}{16\pi^{3}}\frac{1}{4p_{+}p'_{+}}|\mathrsfs M_{ss'\lambda'}|^{2},
\label{eq.spectral.density}
\end{align}
which depends on spin and polarization indices. Averaging over the
spin of the incoming electron and summing over the spin of the outgoing
electron and the polarization of the outgoing photon yields a quantity
which is directly comparable to the classical spectral density
\begin{equation}
\frac{\d^{2}N_{{\rm quantum}}}{\d\omega'\d\Omega}=\frac{1}{2}\sum_{s,s'=1}^{2}\sum_{\lambda'}\frac{\d^{2}N_{ss'\lambda'}}{\d\omega'\d\Omega}.
\label{eq.spectral.density2}
\end{equation}
Now we construct an invariant cross section by dividing Eq.~\eqref{eq.spectral.density2}
by the normalized number of photons $N_{L}$ in the laser pulse, i.e.
\begin{equation}
\frac{\d^{2}\sigma}{\d\omega'\d\Omega}=\frac{1}{N_{L}}\frac{\d^{2}N_{{\rm quantum}}}{\d\omega'\d\Omega}
\end{equation}
with $N_{L}=\intop_{-\infty}^{\infty}\d t\frac{\langle|\mathbf{S}|\rangle}{\omega}$,
where $\mathbf{S}=\mathbf{E\times\mathbf{B}}$ is the Poynting vector
of the laser field, derived from the vector potential \eqref{eq:vector.potential},
yielding
\begin{equation}
N_{L}=\frac{\omega}{2}\frac{a_{0}^{2}m^{2}}{e^{2}}\intop_{-\infty}^{\infty}\d t\, g(t)^{2}
\end{equation}
with $\intop_{-\infty}^{\infty}\d t\cosh^{-2}\phi/\sigma=2\sigma/\omega$
and $\intop_{-\infty}^{\infty}\d t\exp{(-\phi^{2}/2\sigma^{2})^{2}}=\sqrt{\pi}\sigma/\omega$
for the pulse shapes \eqref{eq:def.sech} and \eqref{eq:def.gauss},
respectively. Using this definition, the total cross section is independent
of the pulse shape function $g$ and the pulse length $\sigma$. (A
different definition of the cross section without this property has
been proposed in \cite{Boca}.) This has been checked numerically
for $a_{0}\to0$ by a comparison of $\frac{\d\sigma}{\d\Omega}=\int\d\omega'\frac{\d^{2}\sigma}{\d\omega'\d\Omega}$
with the differential Klein-Nishina cross section \cite{Landau4},
or $\sigma_{{\rm tot}}$ with the total Klein-Nishina cross section.
In particular, in the limit $y_{1}\to0$ we obtain the total Thomson
cross section $\sigma_{T}=665.25\ {\rm mb}$ accurately, as exhibited
in Fig.~\ref{fig.towards.plane.wave} for three different pulse shapes
and pulse lengths.

\section{Discussion}

\label{sect.results}

\subsection{Monochromatic Limit}

\begin{figure}[t]
\noindent \begin{centering}
\includegraphics[angle=270,width=8cm]{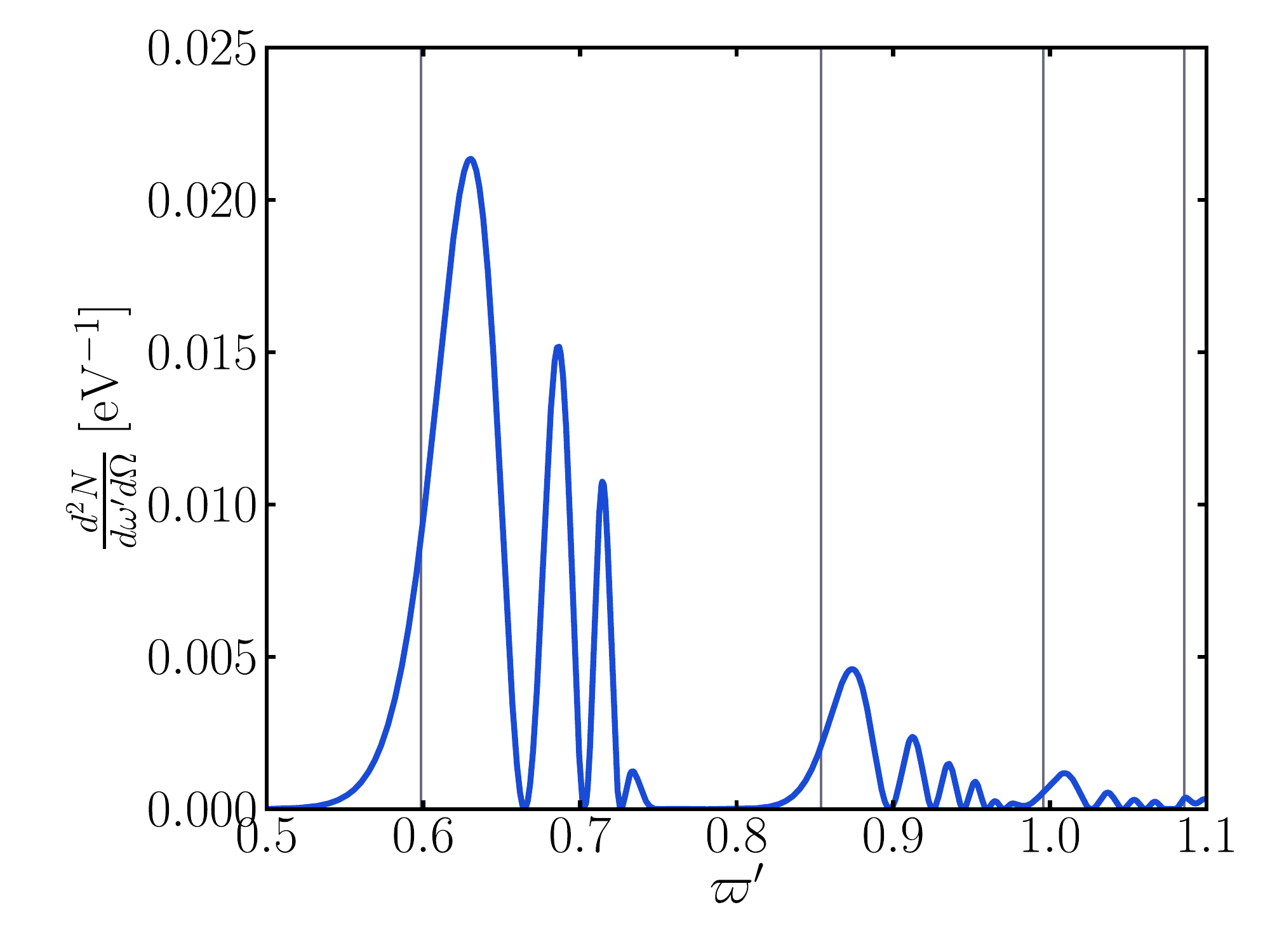}
\includegraphics[angle=270,width=8cm]{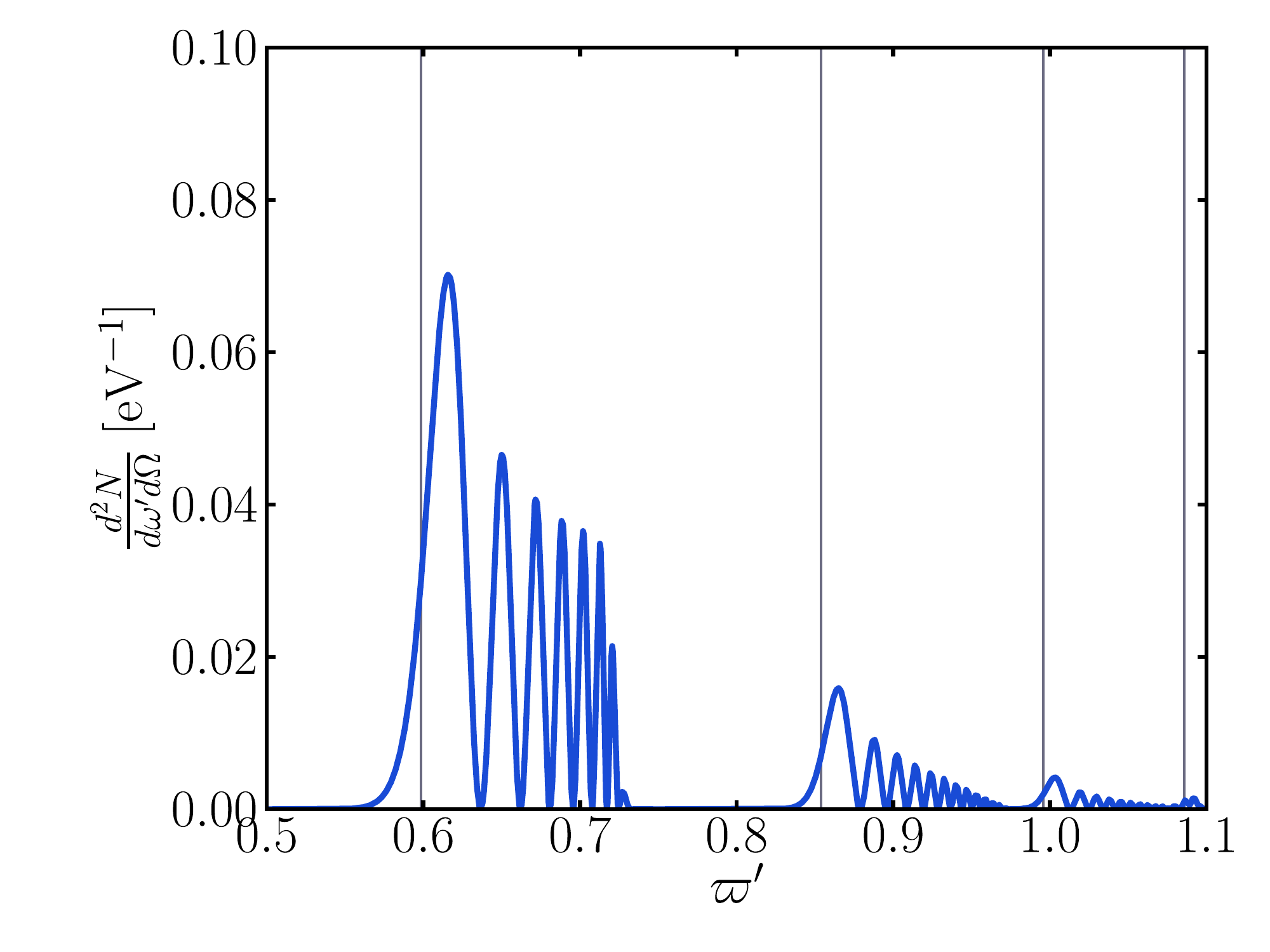}\\
\includegraphics[angle=270,width=8cm]{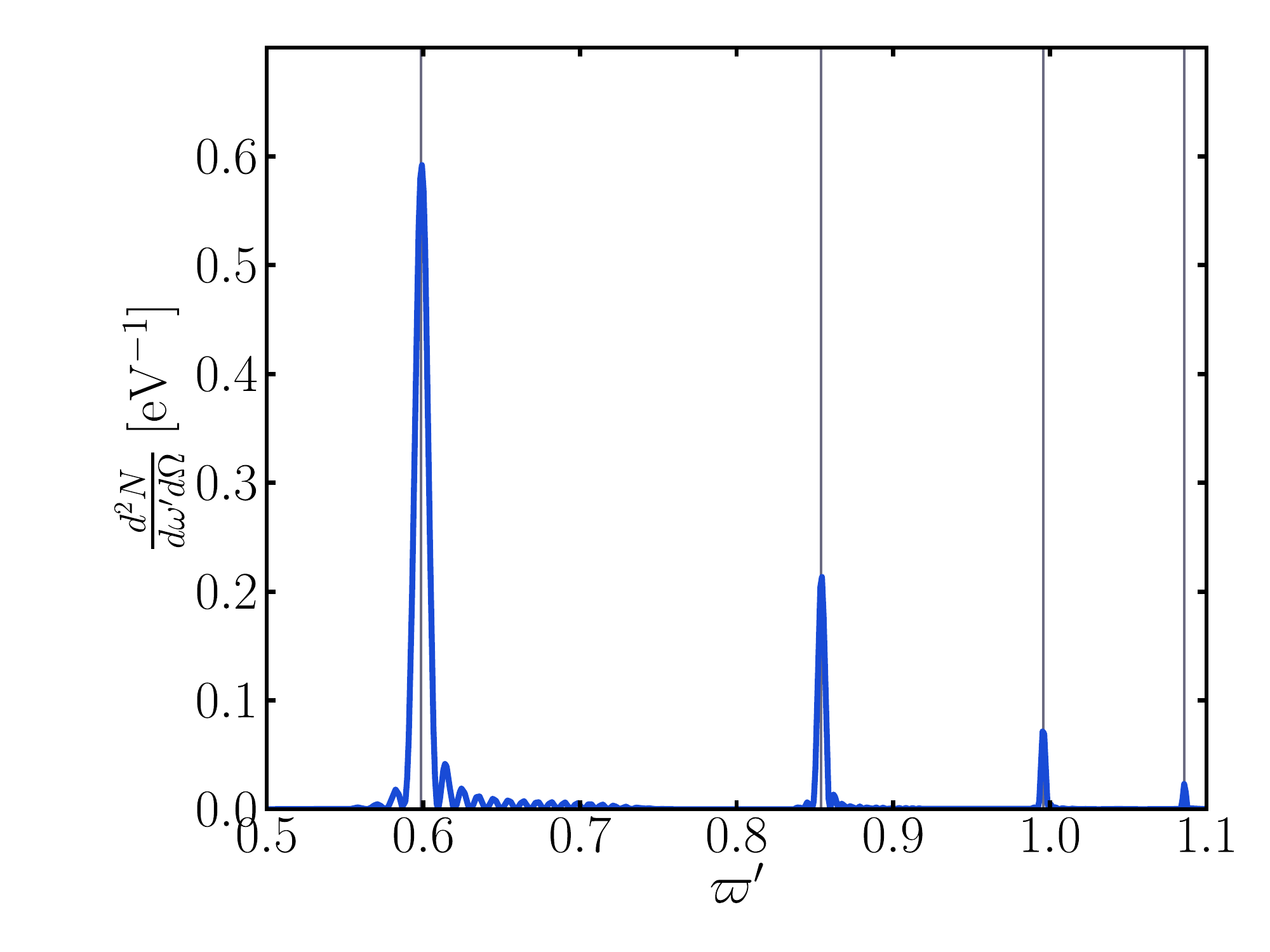}
\par\end{centering}

\caption{Top panels: Spectral density $\d^{2}N/\d\omega'\d\Omega$ as a function
of the scaled frequency $\varpi'=\omega'/\omega'_{1,{\rm classical}}$
for a \emph{cosh} pulse with $\sigma=20$ (left) and $50$ (right).
Bottom panel: Spectral density for a flat-top pulse with \textit{cosh}
edges, $\tau=30$ and $\sigma=20$. In all panels $a_{0}=1.0$,
$\gamma=10^{5}$, $\omega=1.5\ {\rm eV}$, $\theta=1/\gamma$ and
$\varphi=0$. The thin vertical lines depict the non-linear Compton
energies defined in Eq.~\eqref{eq.nlcompton.frequenz}.}

\label{fig.pulse_shape_explanation}
\end{figure}

In the famous case of monochromatic Compton scattering, the frequency
of the scattered photon is uniquely defined by the scattering angle.
For a finite temporal laser pulse, however, this tight relation is
lost. As outlined in subsection \ref{sect.cross.section}, there is
a distribution of the emitted photons for a fixed angle. As an example,
we exhibit in the left top panel of Fig.~\ref{fig.pulse_shape_explanation},
the spectral density $\d^{2}N/\d\omega'\d\Omega$ as a function of
$\varpi'=\omega'/\omega'_{1,{\rm classical}}$ for fixed $\Omega$.
The vertical thin lines depict the positions of the harmonics for
a monochromatic plane wave with infinite duration and the same value
of $a_{0}$, given by \cite{Landau4}
\begin{align}
\omega'_{\ell}=\omega'_{\ell,{\rm quantum}} & =\frac{\ell k\cdot q}{(q+\ell k)\cdot n'},\label{eq.nlcompton.frequenz}
\end{align}
introducing the intensity dependent quasi-momentum of the electron
$q_{-}^{(\prime)}=p_{-}^{(\prime)}+b_{p^{(\prime)}}k_{-}$ with $b_{p^{(\prime)}}=m^{2}a_{0}^{2}/4k\cdot p^{(\prime)}$
and the dressed mass-shell relation $q^{2}=q'^{2}=m_{*}^{2}=m^{2}(1+a_{0}^{2}/2)$.
Note that only $p_-^{(\prime)}$, the conjugate momentum to $x_{+}$, is modified by an intensity
dependent contribution, i.e.~$q_{+}^{(\prime)}=p_{+}^{(\prime)}$
and $\mathbf{q}_{\perp}^{(\prime)}=\mathbf{p}_{\perp}^{(\prime)}$.
The integer $\ell$ labels the individual harmonics, which are not
equidistant in general.

In a pulsed laser field, each harmonic consists of a bunch of spectral
{}``lines'' (or subpeaks) visible in the top panels of Fig.~\ref{fig.pulse_shape_explanation}
with a certain width $\Delta\omega_{\ell}'$ determined by the minimum
and maximum values of intensity in the laser pulse. The high-energy
tail of each harmonic bunch is given by $\omega'_{\ell}(a_{0}\to0)$,
and is produced at the edges of the laser pulse. The low-energy edge
is given by $\omega'_{\ell}(a_{0})$ and accounts for the maximum
red-shift at the center of the pulse. Thus, the spectral width of
each harmonic $\ell$ is given by
\begin{eqnarray}
 & \begin{array}{ccl}
\Delta\omega'_{\ell} & = & \omega'_{\ell}(a_{0}\to0)-\omega'_{\ell}(a_{0})\\
 & = & \omega'_{\ell}(a_{0})\omega'_{\ell}(a_{0}\to0){\displaystyle \frac{b_{p}k\cdot n'}{\ell k\cdot p}}.\end{array}\label{eq.spectral.width}
\end{eqnarray}
The number of subpeaks in a bunch is proportional to the pulse length
$\sigma$ and the intensity $a_{0}^{2}$. The highest subpeak takes
its maximum value at a higher frequency $\omega'$ than predicted
by \eqref{eq.nlcompton.frequenz}, thus, at a smaller intensity-dependent
red-shift than the monochromatic harmonics due to a lower average
$a_{0}$. Hence, one could say that this maximum is blue-shifted as
compared to the monochromatic plane wave.

Increasing $\sigma$ from $20$ to $50$ does not lead to an accumulation
of spectral weight at the non-linear Compton frequencies as could
be expected naively. The number of subpeaks increases but the average
shape of the harmonic bunch is more or less the same for $\sigma=20$
and $50$ with the same spectral width. In fact, to obtain the monochromatic
limit, it is not efficient to take simply the limit $\sigma\to\infty$.
A method with better convergence is to introduce a flat-top area in
the pulse. This however, introduces a second pulse length parameter:
The total pulse length now consists of the rise ''time'' $\sigma$
and the flat-top ''time'' $\tau$. The flat-top part of the pulse
is parametrized as $g_{{\rm flat}}(\phi)=\Theta(\phi+\pi\tau)\Theta(\pi\tau-\phi)$,
where a factor $\pi$ is introduced so that $\tau$ is comparable
to the Gaussian and \emph{cosh} widths $\sigma$ in terms of laser
oscillations under the envelope, and $\Theta(\phi)$ is the Heaviside
step function. Then, the complete pulse is parametrized as
\begin{eqnarray}
g_{{\rm tot}}(\phi;\sigma,\tau) & = & \Theta(\phi+\pi\tau)\Theta(\pi\tau-\phi)+g(\phi-\pi\tau)\Theta(\phi-\pi\tau)+g(\phi+\pi\tau)\Theta(-\phi-\pi\tau).
\end{eqnarray}
The spectrum converges rather fast to sharp peaks centered at the
non-linear Compton frequencies upon increasing $\tau$ from $0$ to
$30$ while keeping $\sigma=20$ constant, as seen in the bottom panel
of Fig.~\ref{fig.pulse_shape_explanation}: The strengths are located
at the sharp non-linear Compton energies. The remaining wiggles around
the non-linear Compton energies vanish upon increasing $\tau$ further. 

In the monochromatic limit $\tau\to\infty$, the rising and trailing
edges of the pulse shape function become unimportant, i.e.~$g\to1$,
and the function $H_{+}$ in \eqref{eq.def.Hplus} reduces to
\begin{eqnarray}
 & \begin{array}{cl}
H_{+}= & \frac{1}{2}(k'_{-}+q'_{-}-q_{-})x_{+}+\alpha_{1}\sin\omega x_{+}-\alpha_{2}\cos\omega x_{+}\\
 & \quad-\frac{b_{p}-b_{p'}}{2}(\cos^{2}\xi-\sin^{2}\xi)\sin2\omega x_{+}\end{array}\label{eq.Hplus.plane}
\end{eqnarray}
with $\alpha_{i}=ma_{0}(\epsilon_{i}\cdot p/k\cdot p-\epsilon_{i}\cdot p'/k\cdot p')$
and re-identifying the electron quasi-momenta $q_{-}^{(\prime)}$.
Upon plugging \eqref{eq.Hplus.plane} into \eqref{eq.matrixelemet.lightcone}
and expanding into a Fourier series, one obtains a fourth energy-momentum
conservation by integrating over $x_{+}$, yielding $q_{-}+\ell k_{-}=q'_{-}+k_{-}^{\prime}$.
The four energy-momentum constraints together lead again to Eq.~\eqref{eq.nlcompton.frequenz}.

The individual harmonics, consisting of a multitude of subpeaks, begin
to overlap if the lower edge of the $\ell+1$st harmonic coincides
with the upper edge of the $\ell$th harmonic, i.e.
\begin{eqnarray}
\omega'_{\ell}(a_{0}\to0) & \geq & \omega'_{\ell+1}(a_{0}).\label{eq.overlap.condition}
\end{eqnarray}
This happens always for sufficiently large values of $a_{0}$ and
$\ell$. The notion of individual harmonics becomes inappropriate,
as one rather observes a continuous spectral distribution.

\subsection{Comparison with Thomson scattering}

\begin{figure}[t]
\noindent \centering{}
\includegraphics[angle=-90,width=8cm]{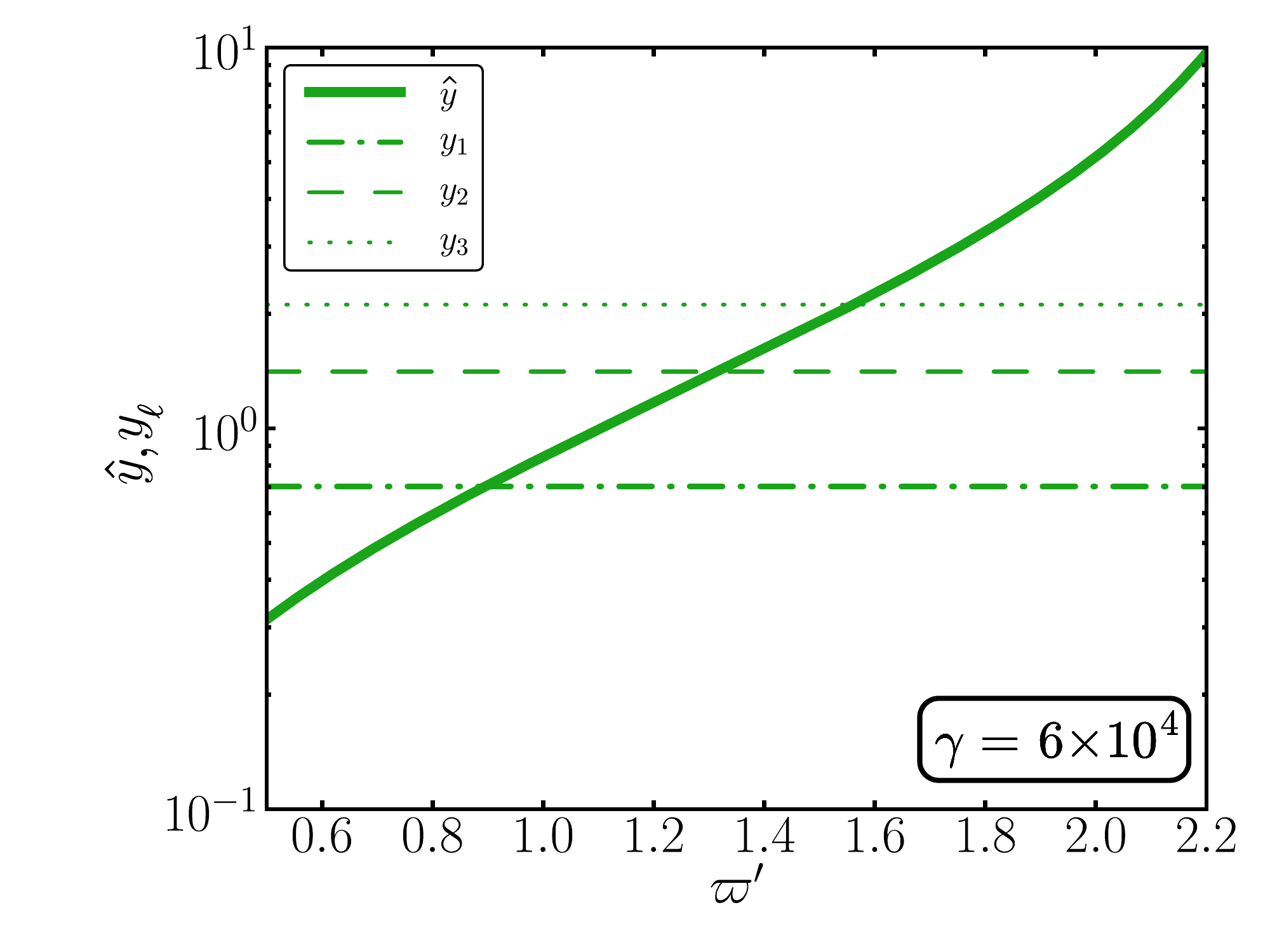}
\caption{Different recoil parameters $\hat{y}(\omega')$ and $y_{\ell}$ for
$\ell=1,2,3$ as a function of the scaled frequency $\varpi'=\omega'/\omega'_{1,{\rm classical}}$.}
\label{fig.y}
\end{figure}

There are different bookkeeping parameters for the characterization
of the Thomson regime as limiting case of the presently considered
scenario. One parameter is $y_{\ell}$ introduced in Eq.~\eqref{eq.def.yl}.
An alternative would be to employ the outgoing momenta instead of
the incoming ones, defining $\hat{y}=(\hat{s}-m^{2})/m^{2}$ with
$\hat{s}=(p'+k')^{2}$. When four-momentum conservation holds, both
definitions coincide (since $k'$ and $p'$ both depend on $\ell$)
and $\hat{s}$ coincides with the usual Mandelstam variable $s$.
However, this is not the case here. These recoil parameters are compared
in Fig.~\ref{fig.y}. The parameter $\hat{y}$ is a function of $\omega'$,
as it depends on $\omega'$ through $k'$
\begin{eqnarray}
\hat{y} & = & 2{\displaystyle \frac{p'\cdot k'}{m^{2}}}=\frac{2}{m^{2}}\frac{(p\cdot k')\,(p\cdot k)}{(p\cdot k-k'\cdot k)}=\frac{2\gamma^{2}\omega'(1-\mathbf{n'\cdot\bbeta})}{m\gamma-\omega'\frac{1-\mathbf{n\cdot n'}}{1-\mathbf{n}\cdot\bbeta}},\label{eq.def.yhat}
\end{eqnarray}
which
 diverges at $\omega_{\infty}^{'}=\frac{m\gamma(1-\mathbf{n\cdot}\bbeta)}{1-\mathbf{n\cdot n'}}$,
defining the boundary of phase space. Thus the physical phase space
is given by $0\leq\theta<\pi$, $0\leq\varphi<2\pi$ and $0<\omega'<\omega'_{\infty}.$
An interpretation of the phase space boundary will be given in subsection
\ref{sub.scaling}.

\begin{figure}[th]
\noindent \begin{centering}
\includegraphics[angle=-90,width=8cm]{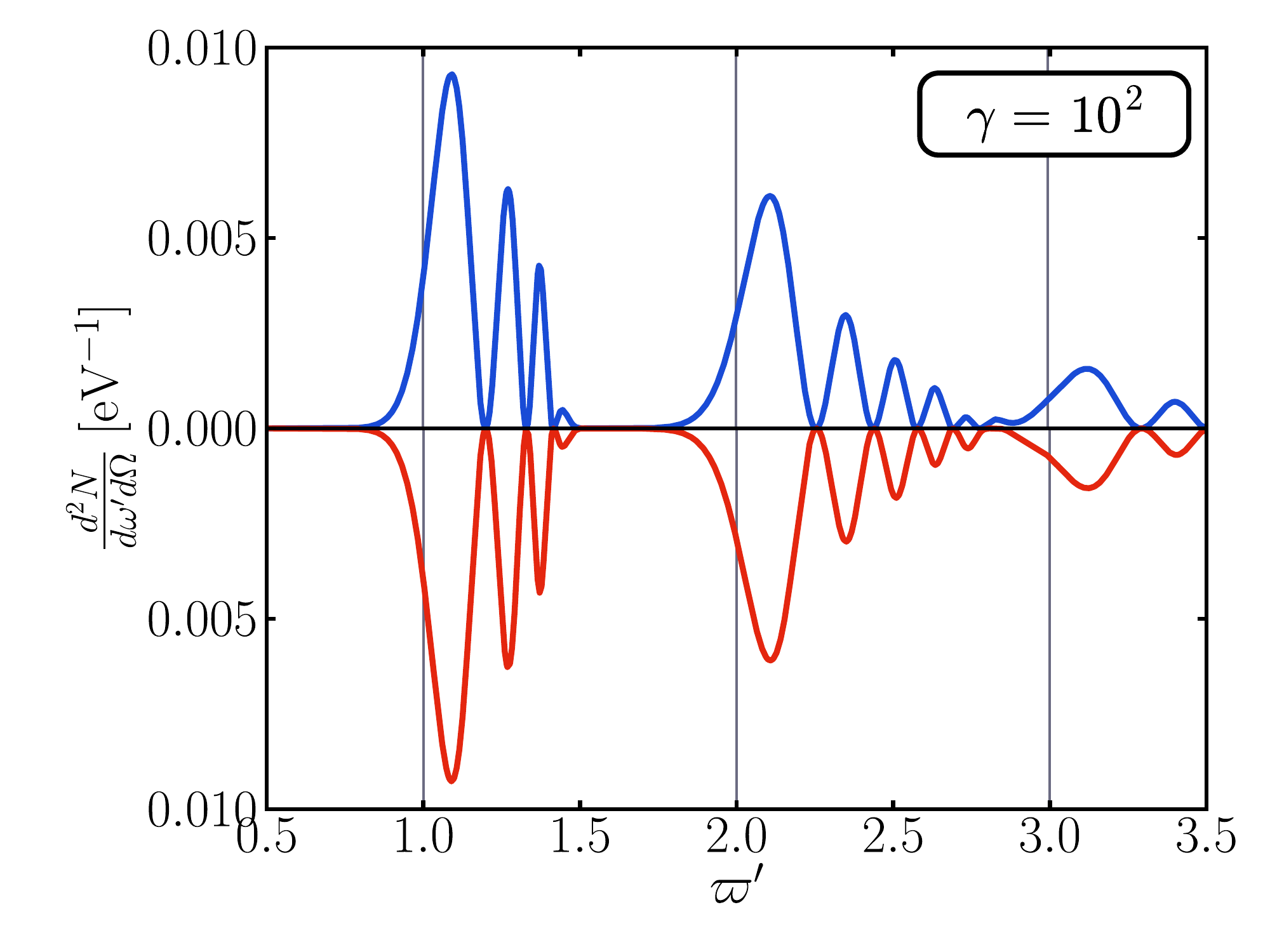} 
\includegraphics[angle=-90,width=8cm]{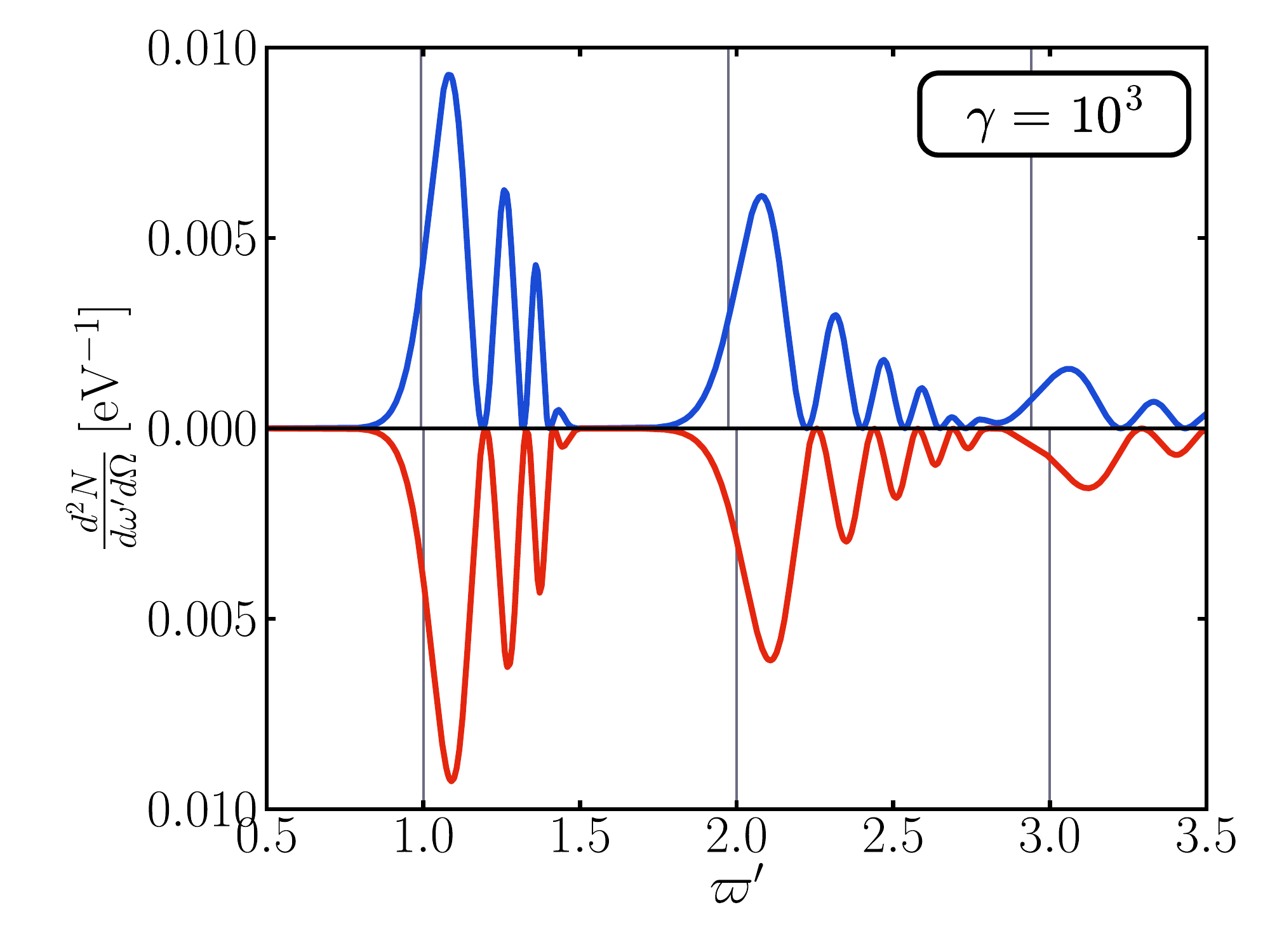} \\
\includegraphics[angle=-90,width=8cm]{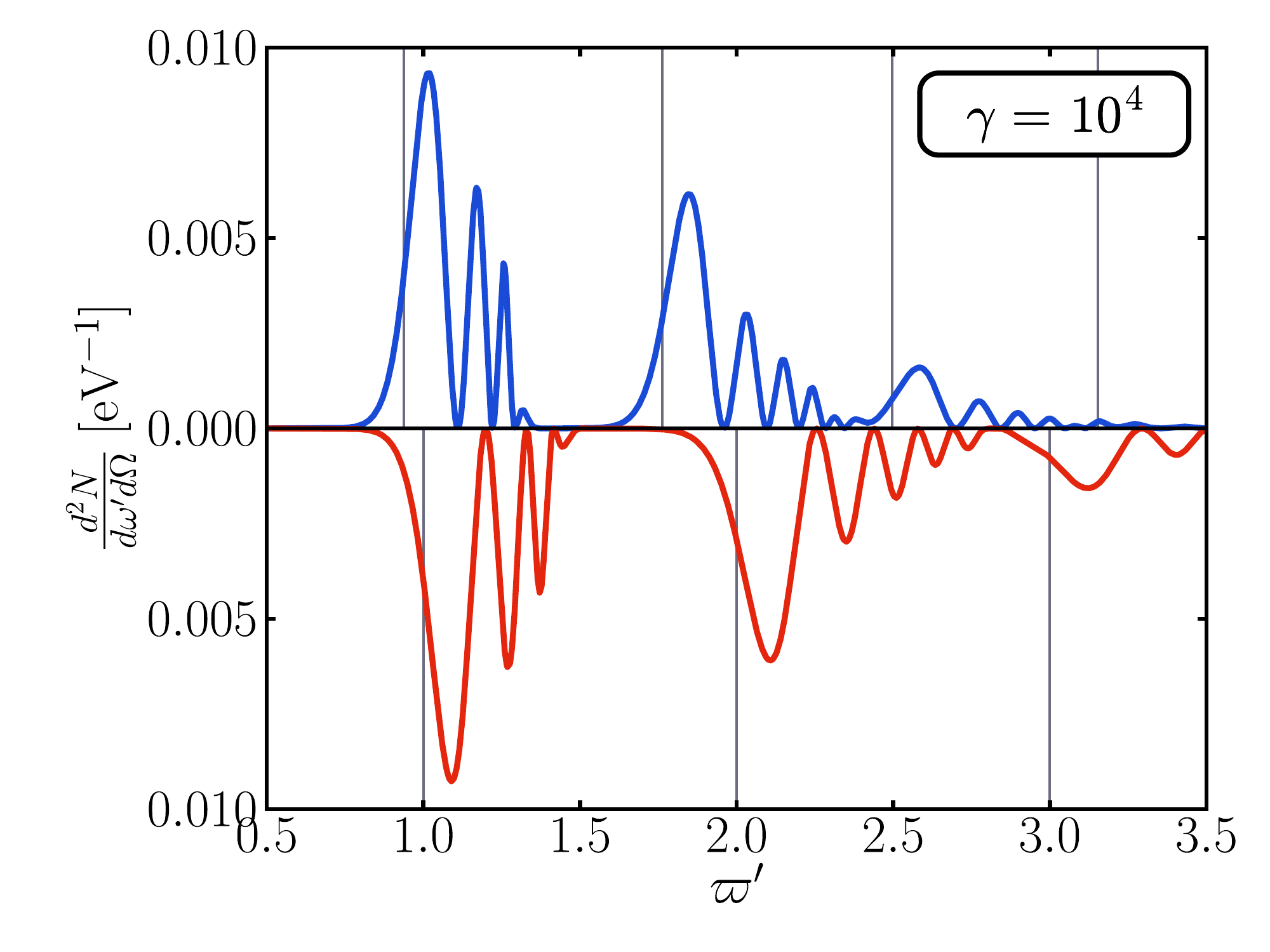} 
\includegraphics[angle=-90,width=8cm]{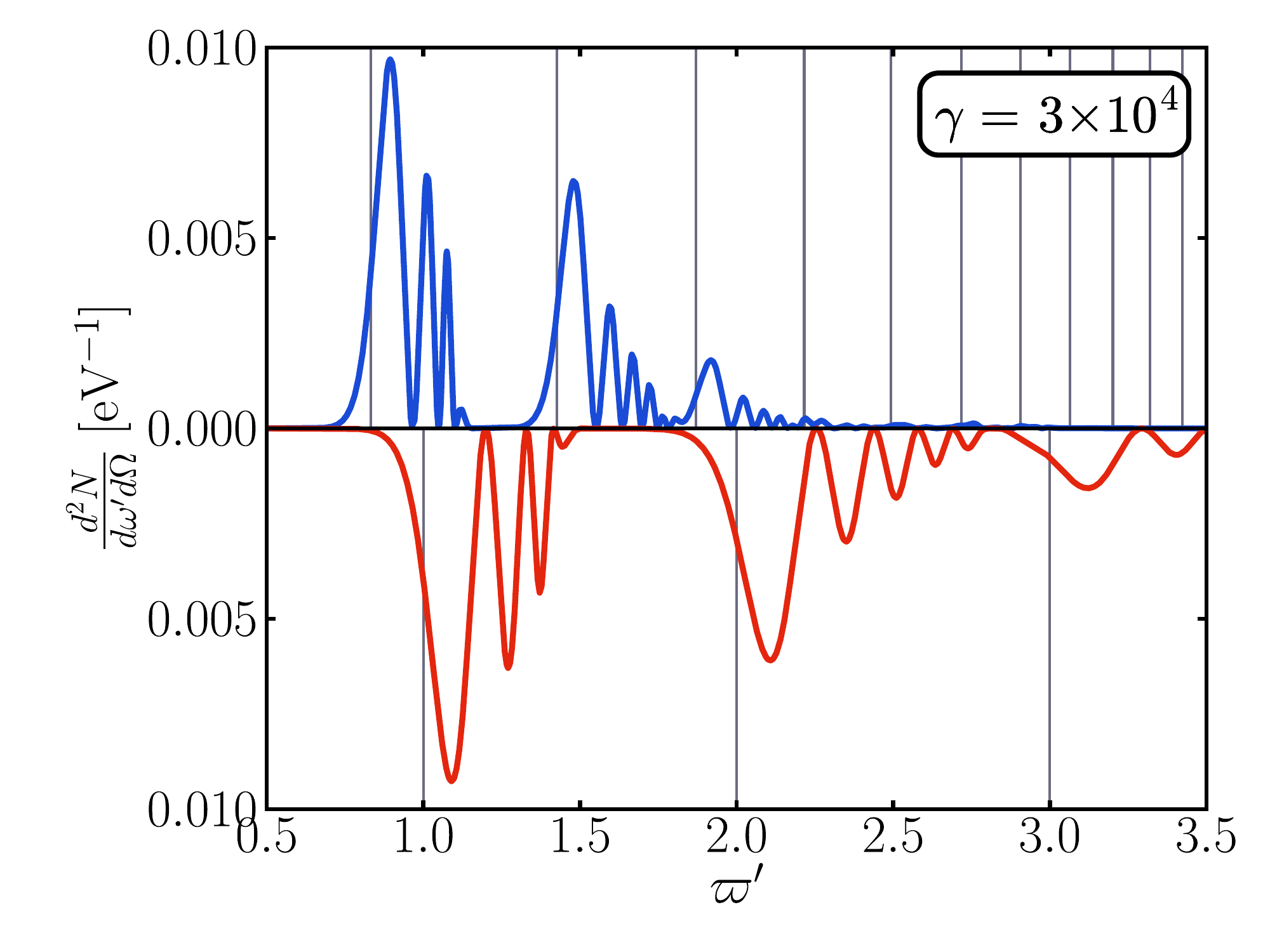} \\
\includegraphics[angle=-90,width=8cm]{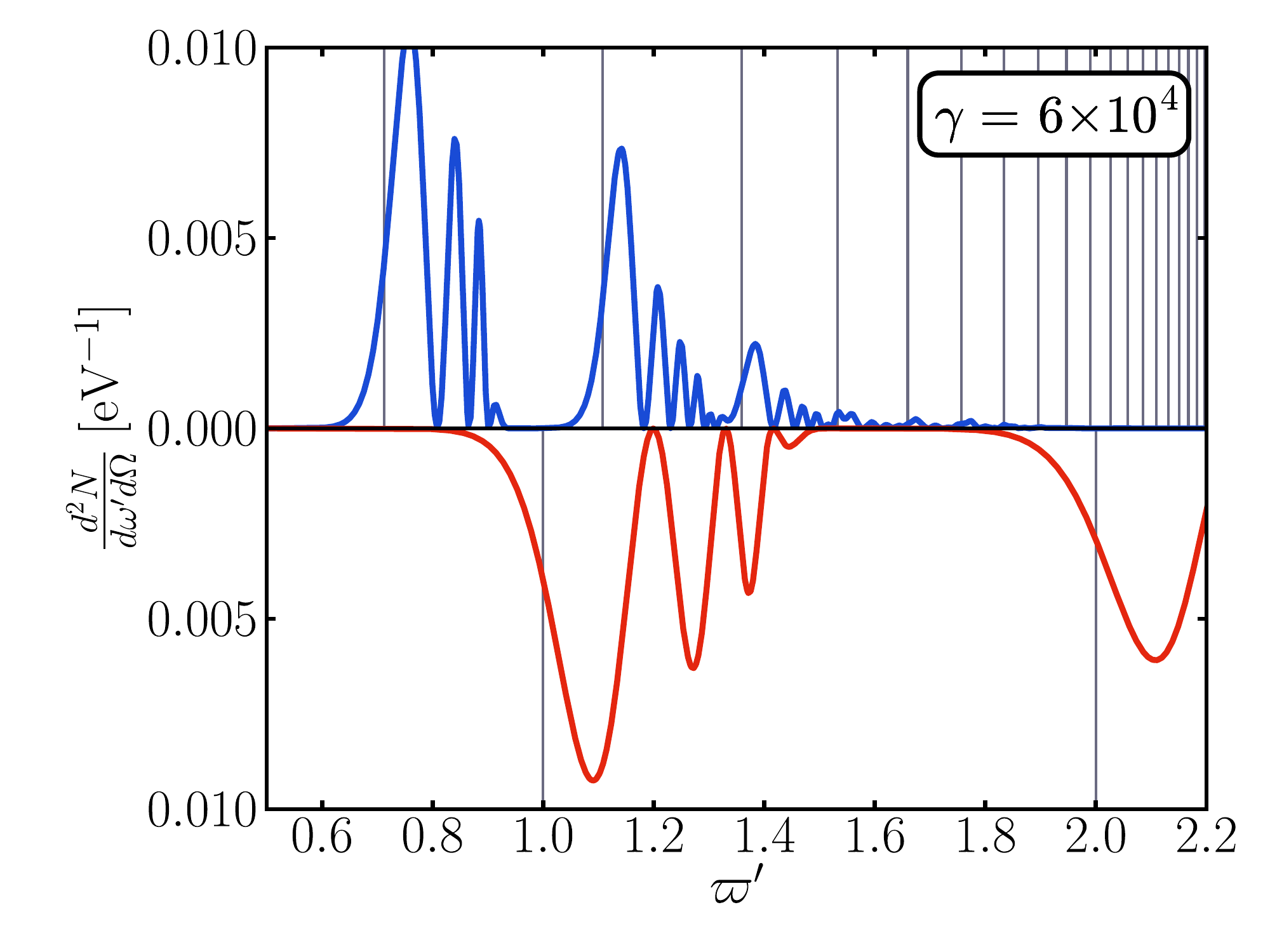} 
\includegraphics[angle=-90,width=8cm]{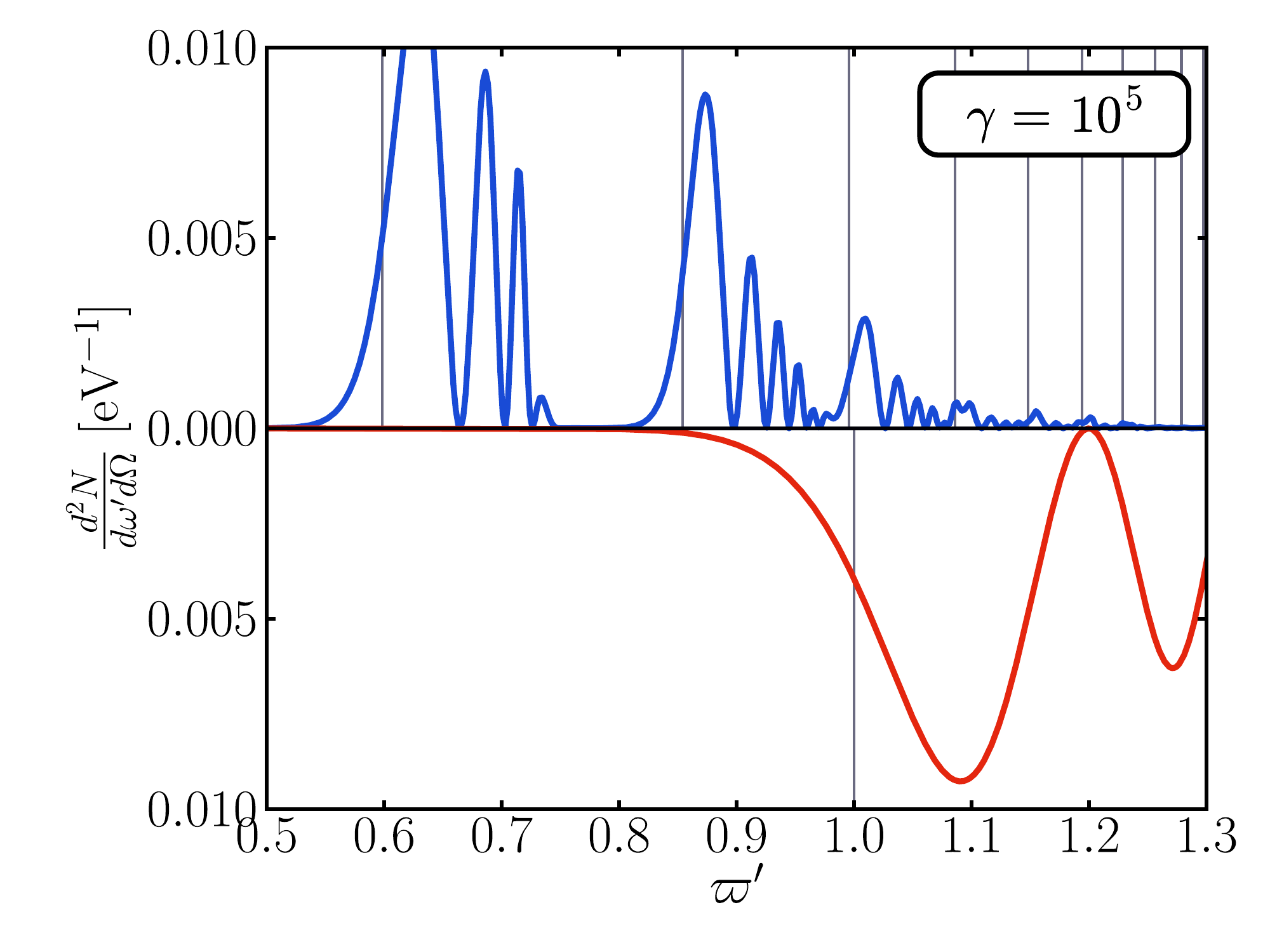}
\par\end{centering}

\caption{The photon spectrum as a function of the scaled frequency $\varpi'=\omega'/\omega'_{1,{\rm {\rm classical}}}(\theta)$
for $\gamma=10^{2},10^{3},10^{4},3\times10^{4},6\times10^{4},10^{5}$
(i.e.~$y_{1}=0.0012,0.012,0.12,0.35,0.7,1.2$) from top left to bottom
right for $\theta=1/2\gamma$ and $\varphi=0$ for a \textit{cosh} pulse shape. The upper (lower) blue (red) curves
are for a quantum (classical) calculation of Compton (Thomson) scattering.
The vertical gray lines mark the positions of the monochromatic harmonics
Eqs.~\eqref{eq.nlcompton.frequenz} and \eqref{eq.classical.harmonic},
respectively. The other parameters are $a_{0}=1.0$, $\omega=1.5\ {\rm eV}$,
$\sigma=20$ and $\xi=0$, i.e.~linear laser polarization.}
\label{fig.thomson.to.compton}

\end{figure}

To relate the Compton amplitude with the classical Thomson counterpart
it is instructive to consider the phase exponential, e.g.~$H_{+}\pm\omega x_{+}$
in $\mathrsfs A_{\pm1}^{1}$, cf.~Eq.~\eqref{eq.def.A_functions}.
For the sake of simplicity, a backscattering head-on geometry with
a circularly polarized laser is assumed in this subsection. Then,
after using some light-cone algebra, the phase reads
\begin{eqnarray}
H_{+}\pm\omega x_{+} & = & \bigg[(k'_{-}+p'_{-}-p_{-})\pm\omega\bigg]x_{+}-\left(\frac{m^{2}a_{0}^{2}}{2k\cdot p}-\frac{m^{2}a_{0}^{2}}{2k\cdot p'}\right)\intop^{\omega x_{+}}\d\phi\, g^{2}(\phi)\\
 & = & \bigg[\frac{k'\cdot p}{n_{+}\cdot p'}\pm\omega\bigg]x_{+}+\frac{k'\cdot k}{(k\cdot p)(k\cdot p')}\frac{m^{2}a_{0}^{2}}{2}\intop^{\omega x_{+}}\d\phi\, g^{2}(\phi).\label{eq.phase.quantum}
\end{eqnarray}
Momentum conservation implies $n_{+}\cdot p'=n_{+}\cdot p-n_{+}\cdot k'$ (see~\eqref{eq.momentum.conservation}).
For $n_{+}\cdot k'\ll n_{+}\cdot p$, the leading term
\begin{eqnarray}
\bigg[\frac{k'\cdot p}{n_{+}\cdot p}x_{+}\pm\omega\bigg]+\frac{k'\cdot k}{(k\cdot p)^{2}}\frac{m^{2}a_{0}^{2}}{2}\intop^{\omega x_{+}}\d\phi\, g^{2}(\phi) & = & k'\cdot x(\tau)\pm\omega x_{+}\label{eq.phase.classical}
\end{eqnarray}
agrees with the corresponding expression obtained in a classical calculation
for Thomson scattering (cf.~\cite{SeiptHeinzl}, see also Eq.~\eqref{eq.retarded.current}).
The frequency of back-scattered photons in monochromatic plane waves
is obtained from \eqref{eq.nlcompton.frequenz} by neglecting $\ell k$
w.r.t. $q$ in the denominator, i.e.
\begin{equation}
\omega'_{\ell,{\rm classical}}=\ell\frac{k\cdot q}{q\cdot n'}.\label{eq.classical.harmonic}
\end{equation}
In Thomson scattering, the harmonics are always equidistant. A series
of plots showing the transition from Thomson to Compton scattering
is exhibited in Fig.~\ref{fig.thomson.to.compton}. The deviations
between Thomson and Compton scattering are: ($i$) a non-linear red-shift
in frequency and ($ii$) a slight modification in the amplitude starting
notably at $y_{1}=0.12$. It is obvious that the red-shift is much
more pronounced at higher frequencies. Figure \ref{fig.thomson.to.compton}
quantifies the well known fact \cite{Landau4} that Compton scattering
turns into Thomson scattering in the low-energy limit. For the chosen
parameters ($a_{0}=1.0,\omega=1.5\ {\rm eV}$) the differences become
significant for $\gamma\geq10^{4}.$ Very drastic differences are
obvious for $\gamma=10^{5}$ (bottom right panel of Fig.~\ref{fig.thomson.to.compton}).

\subsection{Scaling properties of the spectral density}

\label{sub.scaling}

The classical and quantum spectral densities for arbitrary pulse shapes
are connected by the scaling law
\begin{align}
\frac{\d^{2}N_{{\rm classical}}}{\d\omega'\d\Omega}(\omega',\theta) & =\eta\frac{\d^{2}N_{{\rm quantum}}}{\d\omega'\d\Omega}(\chi\omega',\theta)\label{eq:scaling.compton.rate.2}
\end{align}
with the two scaling factors $\eta$ and $\chi$ which are determined
by the monochromatic results. The frequency scaling factor $\chi$
is given by
\begin{align}
\chi & =\frac{\omega'_{\ell,{\rm quantum}}}{\omega'_{\ell,{\rm classical}}}=\frac{n'\cdot u+n'\cdot n\frac{a_{0}^{2}}{4n\cdot u}}{n'\cdot u+n'\cdot n\left(\frac{a_{0}^{2}}{4n\cdot u}+\ell\frac{\omega}{m}\right)}=\frac{q\cdot n'}{(q+\ell k)\cdot n'}.
\end{align}
A continuous effective $\ell_{{\rm {\rm eff}}}$ has to be used, which
follows from the inversion of $\omega'_{\ell,{\rm quantum}}$, yielding
\begin{align}
\ell_{{\rm eff}}(\omega') & =\frac{\frac{\omega'}{\omega}\left(n'\cdot u+n'\cdot n\frac{a_{0}^{2}}{4n\cdot u}\right)}{n\cdot u-n'\cdot n\frac{\omega'}{m}}=\frac{q\cdot k'}{q\cdot k-k'\cdot k},
\end{align}
which simplifies to $\chi=1-k'\cdot k/p\cdot k=1-k_{+}'/p_{+}$. The
scaling of the frequency naturally also includes the scaling behavior
of the phase space factor which is proportional to $\omega'^{2}$.

The scaling factor $\eta$ describes the scaling of the differential
probabilities defined by
\begin{equation}
\eta=\left(\omega_{{\rm quantum}}^{\prime-2}\frac{\d\sigma_{{\rm quantum}}}{\d\Omega}\right)\left(\omega_{{\rm classical}}^{\prime-2}\frac{\d\sigma_{{\rm classical}}}{\d\Omega}\right)^{-1},
\end{equation}
where the differential cross sections $\d\sigma_{{\rm quantum}}$
and $\d\sigma_{{\rm classical}}$ are the monochromatic plane-wave
cross sections, yielding for circular polarization \cite{SeiptHeinzl}
\begin{eqnarray}
\eta &=& \frac{\mathfrak{J}_{\ell}}{\mathfrak{K}_{\ell}}
      = 1 + \frac{x^{2}}{1+x}\, \frac{\mathfrak{L}_{\ell}}{2\mathfrak{L}_{\ell}- \frac{8}{a_{0}^{2}} J_{\ell}^{2}(z)},
\label{eq.scaling.eta}
\end{eqnarray}
where $x=(1-\chi)/\chi$, $y_{*}=2\ell k\cdot p/m_{*}^{2}$ and $\ell_{{\rm eff}}$
has to be used instead of $\ell$ everywhere. The other definitions
are $\mathfrak{L}_{\ell}=J_{\ell+1}^{2}(z)+J_{\ell-1}^{2}(z)-2J_{\ell}^{2}(z)$,
$\mathfrak{K}_{\ell}=-8J_{\ell}^{2}(z)/a_{0}^{2}+2\mathfrak{L}_{\ell}$,
$\mathfrak{J_{\ell}}=\mathfrak{K_{\ell}}+\frac{x^{2}}{1+x}\mathfrak{L_{\ell}}$
and $z=2\ell\sqrt{\frac{a_{0}^{2}/2}{1+a_{0}^{2}/2}}\sqrt{\frac{x}{y_{*}}(1-\frac{x}{y_{*}})}$;
$J_{\ell}$ are Bessel functions of the first kind. In the limit $a_{0}\to0$
one gets
\begin{equation}
\lim_{a_{0}\to0}\eta=1+\frac{x^{2}}{1+x}\frac{1}{2-4\frac{x}{y_{*}}(1-\frac{x}{y_{*}})}
\end{equation}
which is a good approximation for $a_{0}<1$. For linear laser polarization
we expect a similar relation to hold, but with the appropriate linearly
polarized monochromatic plane-wave cross section instead.

\begin{figure}[t]
\noindent \begin{centering}
\includegraphics[height=7.5cm,angle=-90]{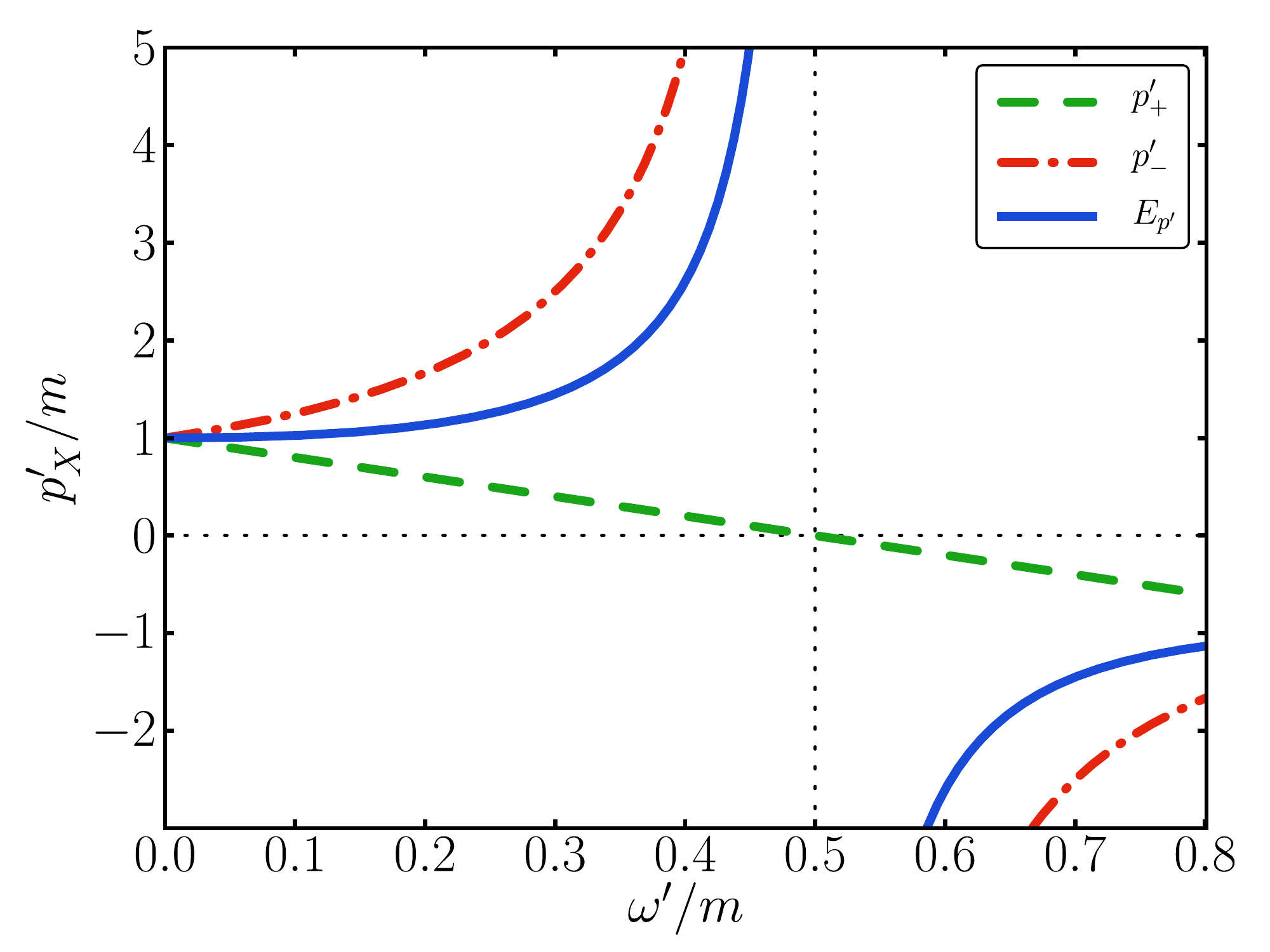}
\includegraphics[height=7.5cm,angle=-90]{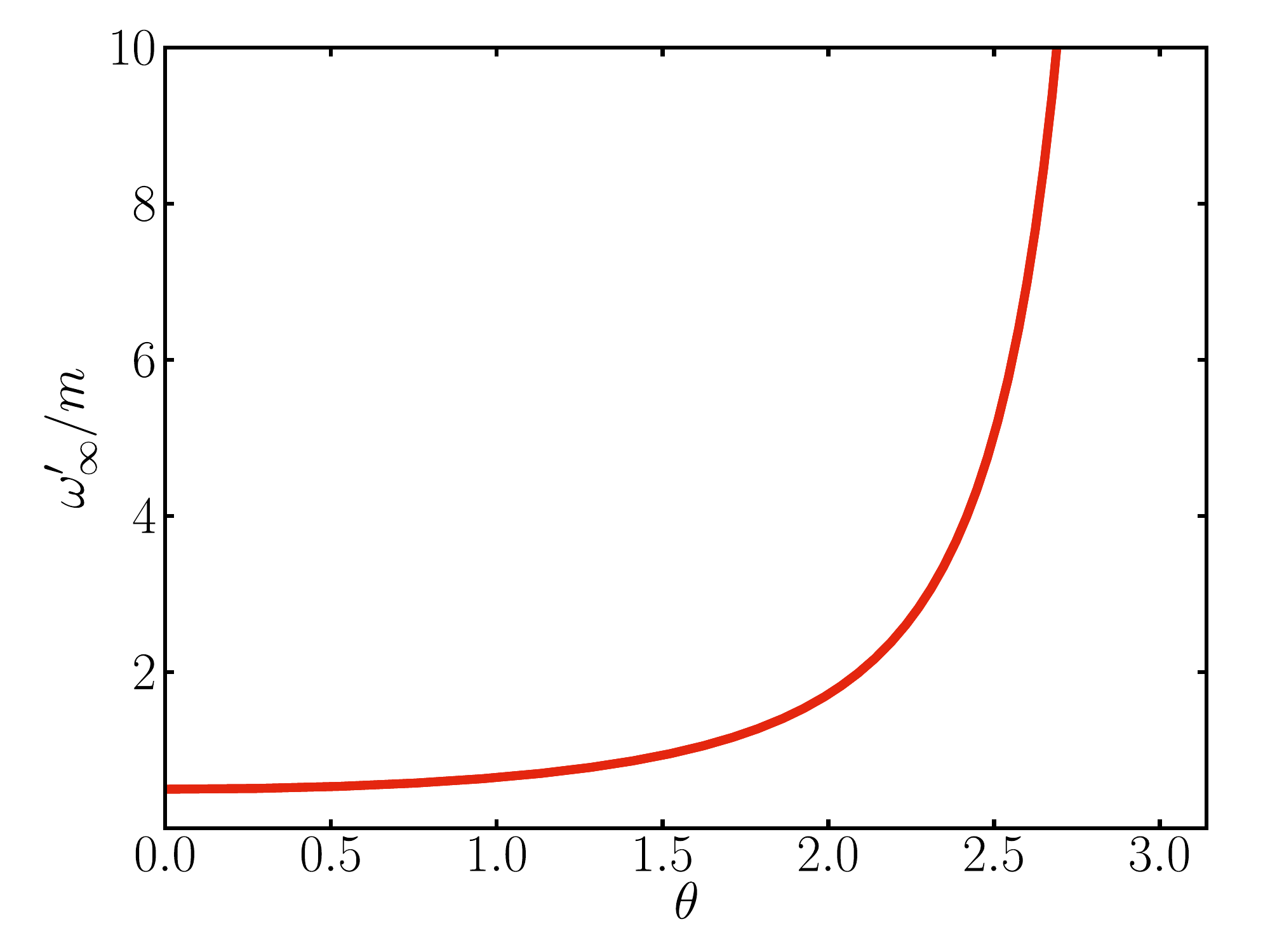}
\par\end{centering}

\caption{Left panel: Different components of the outgoing electron momentum
$p'_{X}$ as a function of frequency $\omega'$ for backscattering
geometry in the rest frame of the incoming electron. Shown are $p'_{-}$
(red, dash-dotted), $p'_{+}$ (green, dashed) and $p'_{0}=E_{p'}$
(blue, solid). The physical phase space has its support at $\chi>0$,
i.e.~for $0<\omega'/m<0.5=\omega_{\infty}'/m$ in this case. Right
panel: Maximum frequency $\omega'_{\infty}$ as function of scattering
angle $\theta$, where $\theta=0$ denotes the backscattering direction.}

\label{fig.phase.space.constraint}
\end{figure}

The scaling function $\chi$ is related to the momentum transfer from
the incoming electron to the outgoing electron via $p'_{+}=\chi p_{+}.$
Thus, $\chi$ is the fraction of $p_{+}$ momentum transferred from
the incoming electron to the outgoing electron $p'_{+}$. Furthermore,
the fraction of momentum transferred to the photon is $k'_{+}=(1-\chi)p_{+}$
thus, $1-\chi$ is another measure of the electron recoil. $\chi$
is a monotonic decreasing function of $\omega'$. The point $\chi(\omega')=0$
corresponds to $p'_{+}=0$ and $k'_{+}=p_{+}$, i.e.~the total amount
of momentum is transferred from the electron to the photon. Further
increasing $\omega'$ would render $\chi$ as well as $p'_{+}$ negative.
This defines the boundary of the physical phase space for the outgoing
particles. When $p'_{+}<0$, then, due to the free particle dispersion
relation \eqref{eq.lightcone.dispersion.relation}, also $p'_{-}<0$.
This, however, would lead to a negative energy $E_{p'}$ due to $E_{p'}=(p'_{+}+p'_{-})/2<0$.
Consequently, there exists a maximum frequency $\omega'_{\infty}$,
defined by $\chi(\omega'_{\infty})=0$:
\begin{equation}
\omega'_{\infty}=\frac{p\cdot k}{n'\cdot k}=\frac{m\gamma(1-\mathbf{n}\cdot\boldsymbol{\beta})}{1-\mathbf{n}\cdot\mathbf{n}'},
\end{equation}
which can also be obtained as the limit ${\displaystyle \lim_{\ell\to\infty}\omega'_{\ell}=\omega'_{\infty}}$,
with $\omega'_{\ell}$ from Eq.~\eqref{eq.nlcompton.frequenz}. This
also coincides with the singularity in $\hat{y}$, see Eq.~\eqref{eq.def.yhat}.
For the backscattering head-on geometry we obtain $\omega'_{\infty}=m/2$
for electrons initially at rest and $\omega'_{\infty}=m\gamma(1+\beta)/2\approx E_{p}$
for ultrarelativistic particles, i.e.~the maximum backscattered frequency
is determined by the energy of the incoming electron. The momentum components
$p'_{+}$, $p'_{-}$ and $E_{p'}$ are depicted in the left panel
of Fig.~\ref{fig.phase.space.constraint} as a function of $\omega'$
in the electron rest frame. The right panel of Fig.~\ref{fig.phase.space.constraint}
shows the dependence of $\omega'_{\infty}$ on the scattering angle
$\theta$. It takes its minimum at the backscattering direction $\theta=0$
and goes to infinity in the limit $\theta\to\pi$, i.e.~forward scattering.

\subsection{Further differences between the classical and QED calculations}

\begin{figure}[t]
\noindent \begin{centering}
\includegraphics[angle=270,origin=c,width=16cm]{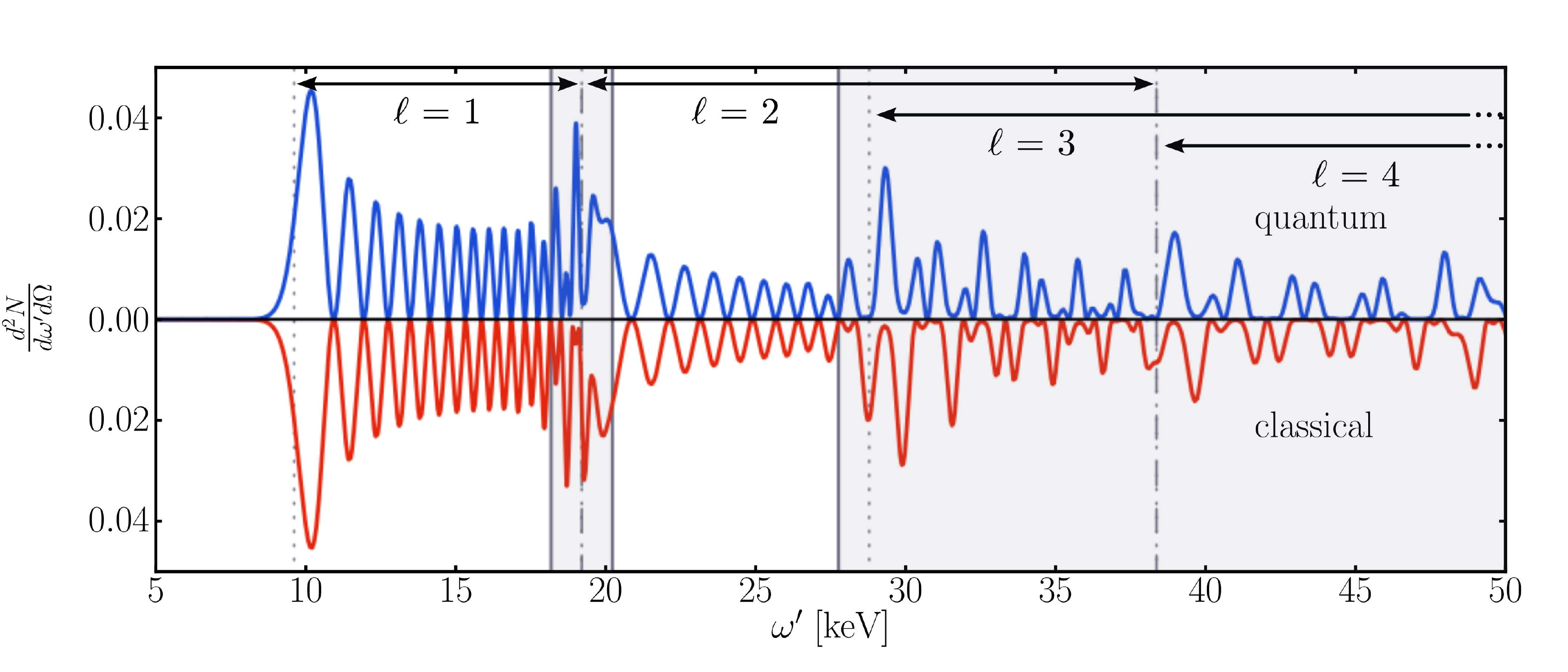}\vspace*{-5cm}
\par\end{centering}

\caption{Comparison of the quantum (upper) and classical (lower) spectral distributions
$\d^{2}N/\d\omega'\d\Omega$ as a function of $\omega'$ for fixed
angles $\theta=1/\gamma$, $\varphi=0$. The overlap regions of different
{}``harmonics'' are highlighted as gray shaded areas. Parameters
are $\sigma=50$, $\omega=1.5\ {\rm eV}$, $a_{0}=2.0$, $\gamma=80$,
i.e.~$\hat{y}\leq2.5\times10^{-3}$.}

\label{fig:different.patterns.qed.classical}
\end{figure}

As demonstrated in the preceding subsections and quantified by the
scaling law, the main difference between the spectral densities of
Thomson and Compton scattering is the proper treatment of the electron
recoil in the latter one. Additionally, there is another regime where
the quantum description goes beyond a classical calculation, even
if $\hat{y}\ll1$, where the total Thomson and Compton cross sections
are equal in leading order. This happens in regions of phase space
where individual harmonics are overlapping (see Eq.~\eqref{eq.overlap.condition}).
There, the subpeaks in the quantum calculation show completely different
patterns in comparison to a classical calculation, see Fig.~\ref{fig:different.patterns.qed.classical}.
Consequently, the scaling law may not be applied where harmonics are
overlapping. For a better orientation, the spectral ranges of the
individual harmonics are marked in Fig.~\ref{fig:different.patterns.qed.classical},
where the lower (upper) edges are depicted by dotted (dashed) lines.
Due to the finite pulse length $\sigma$, the actual spectral distribution
reaches over these edges by $\mathcal{O}(\tilde{\sigma})$, where
$\tilde{\sigma}=\gamma^{2}(1+\beta)^{2}/\sigma$. The gray shaded
areas mark the overlapping regions with a width of $2\tilde{\sigma}$.

The generation of the subpeaks can be described as an interference
effect \cite{SeiptHeinzl}. Thus, when the harmonics overlap, for
a fixed value of $\omega'$ there are contributions from different
harmonics and their interference reacts very sensitive to subtle changes
in the phase of the $\mathrsfs A_{N}^{M}$ functions, see Eqs.~\eqref{eq.phase.quantum}
and \eqref{eq.phase.classical}. The difference in the spectral distributions
looks qualitatively similar to figure 1 of \cite{hartemannprl}, where
the influence of classical radiation reaction force on the spectrum
was studied. The radiation reaction force also provides an electron
recoil in the classical calculation, slightly changing the phases
and leading to a modified spectrum.%

\section{Summary}

In this paper, we discussed non-linear Compton scattering in the Furry
picture and employed light cone coordinates for temporally shaped
laser pulses. We emphasized the structure of the Volkov wavefunctions
in a pulsed laser field. We clearly uncovered the modifications of
the electron wavefunction due to their interaction with the laser field.
The $S$ matrix element for non-linear Compton scattering was evaluated
in the framework of Volkov states within the Furry picture. An expression
for the cross section which is independent of the pulse shape and
pulse length was presented.

We focused on the differences between classical calculations of non-linear
Thomson scattering and quantum calculations of non-linear Compton
scattering. In both cases, spectral broadening and harmonic substructures
have been found, which are, however, shifted in the quantum calculation.
These harmonic substructures, still lacking an experimental verification,
are interpreted as an interference effect. We found that for $y_\ell \ll1$
the differential quantum transition probability in many cases coincides
with the classical Thomson scattering result also for pulsed laser
fields, i.e.~quantum effects are mostly negligible in this regime.
Also the total Compton cross section coincides with the Thomson cross
section $\sigma_{T}$ for $y_{\ell}<10^{-2}$. 

As a main result, we presented a scaling law, connecting the classical
and quantum spectral densities for arbitrary $y_{\ell}$, e.g.~$y_{\ell}>10^{-2}$,
relating the classical and quantum results. The remarkable feature
is, that the substructures of individual harmonics are also simply
scaled. One might speculate that the scaling law may also be applied
for arbitrary laser beams, in particular, for strongly focused beams.
Hence, it might serve as a tool to add recoil effects to results
obtained within classical Thomson scattering models.

Furthermore, we also found regions in phase space, where the differential
probabilities for Thomson and Compton are different, although $y_{\ell}<10^{-2}$
and the total cross sections coincide. This happens for sufficiently
large values of $a_0$, when the individual harmonics are overlapping.
In these regions of phase space, both spectral densities show different,
almost erratic behavior. This observation is in qualitative agreement
with previous studies of the effect of the radiation reaction force
on the spectrum of non-linear Thomson scattering. The radiation reaction
force introduces an electron recoil in Thomson scattering to the classical
picture. Of course, the scaling law is not applicable in the regions
where the spectral densities show this erratic behavior.
\section*{Acknowledgments}
The authors gratefully acknowledge stimulating discussions with T.
E. Cowan and K. Chouffani.
 \newpage
\appendix

\section{Light cone coordinates}

\label{app.lightcone.coordinates}

\subsection{Basic definitions}

The light cone components of a four-vector $x^{\mu}=(t,x,y,z)$ are
defined as $x_{\pm},\mathbf{x_{\perp}}$ by
\begin{align}
x_{-}=t-z,\qquad x_{+}=t+z,\qquad\mathbf{x}_{\perp}=(x,y).
\end{align}
In these new coordinates, the scalar product reads
$x\cdot y=\frac{1}{2}\left(x_{+}y_{-}+x_{-}y_{+}\right)-\mathbf{x}_{\perp}\cdot\mathbf{y}_{\perp}$.
Arranging these components as a four-vector $X^{\mu}=(x_{+},x_{-},\mathbf{x_{\perp}})$,
one may introduce a non-diagonal metric
\begin{eqnarray}
g_{\mu\nu} & {\displaystyle \ =\ } & \left(\begin{tabular}{cccc}
 \mbox{\ 0\ }  &  \mbox{\ \ensuremath{\frac{1}{2}}\ }  &  \mbox{\ 0\ }  &  \mbox{\ 0\ } \\
\ensuremath{\frac{1}{2}} &  0  &  0  &  0 \\
0  &  0  &  -1  &  0 \\
0  &  0  &  0  &  -1 \end{tabular}\right)\label{eq.def.metric}
\end{eqnarray}
with $\sqrt{-g}=1/2$, thus $\d^{4}x\to\frac{1}{2}\d x_{+}\d x_{-}\d^{2}\mathbf{x}_{\perp}$.
The inverse transformation is given by $x_{0}=\frac{1}{2}(x_{+}+x_{-}),\ x_{3}=\frac{1}{2}(x_{+}-x_{-}).$

More generally, light cone coordinates can be introduced by projecting
onto a light-like four-vector, which is in our case given by the laser wave
vector $k^{\mu}=\omega n_{+}^{\mu}=\omega(1,\mathbf{n})$ defining
a direction $\mathbf{n}$ with $\mathbf{n}\cdot\mathbf{n}=1$. The
coordinate system will always be aligned such that $\mathbf{n}=(0,0,-1)$.
Then the light cone components of a vector $B^{\mu}$ may be defined
by projection via
\begin{align}
B_{+} & =B\cdot n_{+},\quad B_{-}=B\cdot n_{-},\\
\mathbf{B}_{\perp} & =\mathbf{B}-(\mathbf{B}\cdot\mathbf{n})\mathbf{n}
\end{align}
with $n_{-}^{\mu}=(1,-\mathbf{n})$. Additionally, we need the orthonormal
transverse basis vectors $\epsilon_{i},i=(1,2)$ with $\epsilon_{i}\cdot n_{\pm}=0$
and $\epsilon_{i}\cdot\epsilon_{j}=-\delta_{ij}$, i.e.~the transverse
part of $B^{\mu}$ may also be defined as $\mathbf{B}_{\perp}=(B_{1},B_{2}),\ B_{i}=B\cdot\epsilon_{i}$.
With this convention of light cone coordinates, the conjugate momentum
to $x_{+}$ is $P_{-}$ and vice versa.

\subsection{Kinematics in light cone coordinates}

Special light cone coordinates may be defined with respect to the
laser momentum $k^{\mu}=\omega(1,\mathbf{n})$ with $\mathbf{n}=-\mathbf{e}_{z}$.
Let be the momentum of the incoming electron $p^{\mu}=(E_{p},\mathbf{p})=m\gamma(1,\bbeta)=mu^{\mu}$
with $\bbeta=-\beta\mathbf{n}$ and the momentum of the outgoing photon
$k'^{\mu}=(\omega',\mathbf{k}')=\omega'(1,\mathbf{n}')=\omega'n'^{\mu}$
with $\mathbf{n}'=(\cos\varphi\sin\theta,\sin\varphi\sin\theta,\cos\theta)$.
Then the light cone components of these vectors read
\begin{eqnarray}
\begin{array}{c}
k_{-}=2\omega,\\
k'_{-}=\omega'(1+\cos\theta),\\
p_{-}=m\gamma(1-\beta),\end{array} & \qquad\begin{array}{c}
k_{+}=0,\\
k'_{+}=\omega'(1-\mathbf{\cos\theta}),\\
p_{+}=m\gamma(1+\beta),\end{array} & \qquad\begin{array}{c}
\mathbf{k}_{\perp}=0,\\
\mathbf{k}'_{\perp}=\omega'\sin\theta(\cos\varphi,\sin\varphi),\\
\mathbf{p}_{\perp}=0.\end{array}
\end{eqnarray}
The free particle dispersion relation $E_{p}^{2}=\mathbf{p}^{2}+m^{2}$
reads in light cone coordinates
\begin{equation}
p_{-}=\frac{\mathbf{p}_{\perp}^{2}+m^{2}}{p_{+}}.\label{eq.lightcone.dispersion.relation}
\end{equation}
Using momentum conservation, one obtains for the components of $p'$
\begin{align}
\begin{array}{cc}
\mathbf{p'}_{\perp} & =\mathbf{p}_{\perp}-\mathbf{k}'_{\perp},\\
p'_{+} & =p_{+}-k'_{+}.\end{array} \label{eq.momentum.conservation}
\end{align}
It is worth noting that there is in general no conservation law for
the component $p'_{-}$. That component is fixed by the dispersion
relation \eqref{eq.lightcone.dispersion.relation} yielding
\begin{align}
p'_{-} & =\frac{m^{2}+\mathbf{(}p'_{\perp})^{2}}{p'_{+}}=\frac{m^{2}+(\mathbf{p_{\perp}}-\mathbf{k_{\perp}}')^{2}}{p_{+}-k'_{+}}.\label{eq:def.pminus.onshell.constraint}
\end{align}
All the components of $p'$ have to be considered as a function of
$\omega'$ and $\mathbf{n}'$. The projection of $p'$ onto $k$ may
be rewritten as
\begin{align}
k\cdot p'=k_{-}p'_{+}=k_{-}(p_{+}-k'_{+})=k\cdot p-k\cdot k',
\end{align}
from which the important relation
\begin{align}
\frac{1}{k\cdot p}-\frac{1}{k\cdot p'}=-\frac{k\cdot k'}{k\cdot p\ k\cdot p'}
\end{align}
may be derived. Eventually, we present a useful collection of relations
between momentum components:
\begin{eqnarray}
k'_{-}+p'_{-}-p_{-} & = & \frac{2p\cdot k'}{p'_{+}}=2\frac{p\cdot k'}{p'\cdot k},\\
y & = & \frac{2p'\cdot k'}{m^{2}}=\frac{p_{+}(k'_{-}+p'_{-}-p_{-})}{m^{2}}=\frac{p_{+}}{p'_{+}}\frac{2p\cdot k'}{m^{2}},\\
\frac{p'\cdot k'}{p\cdot k'} & = & \frac{p_{+}}{p'_{+}}=\frac{p\cdot k}{p'\cdot k},\\
\chi & = & 1-\frac{k'\cdot k}{p\cdot k}=\frac{p\cdot k'}{p'\cdot k'}.
\end{eqnarray}

\section{Fourier expansion of the Volkov wavefunction}

\label{app.fourier} Here, we provide a Fourier expansion of the Volkov
state \eqref{eq:volkov.state} in a pulsed laser field to gain further
insight to the structure of the Volkov state and to use it as an alternative
way to obtain the matrix element for non-linear Compton scattering
as a convolution in momentum space. We define the Fourier transformations
in $x_{-},x_{+}$ and $\mathbf{x_{\perp}}$ directions with different
signs according to the metric \eqref{eq.def.metric}
\begin{equation}
\begin{array}{cc}
\tilde{F}(\mathbf{Q}_{\perp},Q_{+},Q_{-}) & =\frac{1}{2}\intop\d^{2}\mathbf{x}_{\perp}\d x_{-}\d x_{+}F(\mathbf{x}_{\perp},x_{-},x_{+})e^{\frac{i}{2}(Q_{-}x_{+}+Q_{+}x_{-})-i\mathbf{Q_{\perp}\cdot x_{\perp}}},\\
F(\mathbf{x_{\perp}},x_{-},x_{+}) & =\frac{1}{2}\intop\frac{\d^{2}\mathbf{Q}_{\perp}\d Q_{+}\d Q_{-}}{(2\pi)^{4}}\tilde{F}(\mathbf{Q_{\perp}},Q_{+},Q_{-})e^{-\frac{i}{2}(Q_{-}x_{+}+Q_{+}x_{-})+i\mathbf{Q_{\perp}\cdot x_{\perp}}}.\end{array}
\end{equation}
The Fourier transform of the Volkov matrix $C_{p}(x)$ (see Eq.~\eqref{eq.def.B_p})
reads
\begin{align}
\begin{array}{ccl}
\tilde{C}_{p}(\mathbf{Q}_{\perp},Q_{+},Q_{-}) & = & (2\pi)^{3}\delta^{2}(\mathbf{Q}_{\perp}-\mathbf{p_{\perp}})\delta(Q_{+}-p_{+})\\
 &  & \times\Big\{\G_{0}(Q_{-})+d_{p}\slashed k\big[\slashed\epsilon_{-}\G_{1}(Q_{-})+\slashed\epsilon_{+}\G_{-1}(Q_{-})\big]\Big\}\end{array}
\end{align}
with
\begin{align}
\G_{N}(Q_{-}) & =\int\d x_{+}e^{\frac{i}{2}(Q_{-}-p_{-})x_{+}}g^{|N|}(x_{+})e^{iN\omega x_{+}+if(x_{+})}.
\end{align}
The functions $\G_{N}$ describe the nontrivial, pulse dependent momentum
distribution of the Volkov wavefunction. Using the Fourier representation
of the Volkov state, the matrix element for non-linear Compton scattering
\eqref{eq:matrix.element} reads
\begin{align}
\begin{array}{ccl}
S_{fi} & = & N_{0}(2\pi)^{3}\delta^{2}({\bf p'_{\perp}+k'_{\perp}-p_{\perp}})\delta(p'_{+}+k'_{+}-p_{+})\times\\
 &  & \qquad\times\Big[\mathrsfs T_{0}^{0}(\G_{0}^{*}\star\G_{0})+\mathrsfs T_{-1}^{1}(\G_{0}^{*}\star\G_{1}+\G_{-1}^{*}\star\G_{0})\\
 &  & \qquad\quad+\mathrsfs T_{1}^{1}(\G_{0}^{*}\star\G_{-1}+\G_{1}^{*}\star\G_{0})+\mathrsfs T_{0}^{2}(\G_{1}^{*}\star\G_{1}+\G_{-1}^{*}\star\G_{-1})\\
 &  & \qquad\quad+\mathrsfs T_{-2}^{2}(\G_{-1}^{*}\star\G_{1})+\mathrsfs T_{2}^{2}(\G_{1}^{*}\star\G_{-1})\Big]\end{array}\label{eq.matrix.element.fourier}
\end{align}
with the convolution
\begin{align}
(\G_{n}^{*}\star\G_{m})(k'_{-})\equiv\int\frac{\d Q_{-}}{2\pi}\G_{n}^{*}(k'_{-}-Q_{-})\G_{m}(Q_{-}).\label{eq.convolution}
\end{align}
Comparing \eqref{eq.matrix.element.fourier} with \eqref{eq.S_matrix.final}
and \eqref{eq.def.M_matrix} we find
\begin{equation}
\begin{aligned}\mathrsfs A_{0}^{0} & =\G_{0}^{*}\star\G_{0}, & \qquad\mathrsfs A_{\pm1}^{1} & =\G_{0}^{*}\star\G_{\mp1}+\G_{\pm1}^{*}\star\G_{0},\\
\mathrsfs A_{0}^{2} & =\G_{1}^{*}\star\G_{1}+\G_{-1}^{*}\star\G_{-1}, & \qquad\mathrsfs A_{\pm2}^{2} & =\G_{\pm1}^{*}\star\G_{\mp1}.\end{aligned}
\end{equation}
The representation \eqref{eq.matrix.element.fourier} of the Compton
amplitude together with \eqref{eq.convolution} furnishes another
interpretation of the subpeaks in the Compton rate as the overlap
of the momentum distributions of the incoming and outgoing Volkov
states.

\end{document}